\documentclass[journal]{IEEEtran}
\usepackage{xcolor,soul,framed} 
\usepackage{algorithm}
\usepackage{algpseudocode}
\colorlet{shadecolor}{yellow}
\usepackage[pdftex]{graphicx}
\graphicspath{{../pdf/}{../jpeg/}}
\DeclareGraphicsExtensions{.pdf,.jpeg,.png}

\usepackage[cmex10]{amsmath}
\usepackage{array}
\newcommand{\PreserveBackslash}[1]{\let\temp=\\#1\let\\=\temp}
\newcolumntype{C}[1]{>{\PreserveBackslash\centering}p{#1}}
\newcolumntype{R}[1]{>{\PreserveBackslash\raggedleft}p{#1}}
\newcolumntype{L}[1]{>{\PreserveBackslash\raggedright}p{#1}}

\usepackage{mdwmath}
\usepackage{mdwtab}
\usepackage{eqparbox}
\usepackage{url}
\usepackage{hyperref}
\usepackage{amssymb}

\usepackage{tikz,xcolor,hyperref}
\definecolor{lime}{HTML}{A6CE39}
\DeclareRobustCommand{\orcidicon}{%
	\begin{tikzpicture}
	\draw[lime, fill=lime] (0,0)
	circle [radius=0.16]
	node[white] {{\fontfamily{qag}\selectfont \tiny ID}};	\draw[white, fill=white] (-0.0625,0.095)
	circle [radius=0.007];	\end{tikzpicture}
	\hspace{-2mm}}
\foreach \x in {A, ..., Z}{%
	\expandafter\xdef\csname orcid\x\endcsname{\noexpand\href{https://orcid.org/\csname orcidauthor\x\endcsname}{\noexpand\orcidicon}}
	}

\ifCLASSINFOpdf
\else
\fi
\begin{document}
\title{Voxel Structure-based Mesh Reconstruction from a 3D Point Cloud}

\author{Chenlei~Lv\orcidA{},~\IEEEmembership{Member,~IEEE,}
Weisi Lin\orcidB{},~\IEEEmembership{Fellow,~IEEE,} and~Baoquan Zhao\orcidC{}
\thanks{
This work was supported by the Ministry of Education, Singapore,
under its Tier-2 Fund MOE2016-T2-2-057(S). The authors are with the School of Computer Science and Engineering,
Nanyang Technology University, 639798, Singapore.
(Corresponding author: Weisi Lin.).}}

\markboth{}%
{Shell \MakeLowercase{\textit{et al.}}: Bare Demo of IEEEtran.cls for IEEE Journals}

\maketitle

\begin{abstract}
Mesh reconstruction from a 3D point cloud is an important topic in the fields of computer graphic, computer vision, and multimedia analysis. In this paper, we propose a voxel structure-based mesh reconstruction framework. It provides the
intrinsic metric to improve the accuracy of local region detection. Based on the detected local regions, an initial reconstructed mesh can be obtained. With the mesh optimization in our framework, the initial reconstructed mesh is optimized into an isotropic one with the important geometric features such as external and internal edges. The experimental results indicate that our framework shows great advantages over peer ones in terms of mesh quality, geometric feature keeping, and processing speed. The source code of the proposed method is publicly available\footnote{\href{https://github.com/vvvwo/Parallel-Structure-for-Meshing}{Code Link: github.com/vvvwo/Parallel-Structure-for-Meshing}.}.
\end{abstract}

\begin{IEEEkeywords}
mesh reconstruction, intrinsic metric, voxel structure, isotropic property.
\end{IEEEkeywords}

\IEEEpeerreviewmaketitle

\section{Introduction}

As one kind of 3D object representation, polygonal mesh has been widely used in different applications such as 3D object animation \cite{agudo2018shape}, simplification \cite{cheng2006adaptive}, compression \cite{ahn2012efficient}, digital watermarking \cite{corsini2007watermarked}\cite{jiang2017reversible}, saliency detection \cite{jeong2017saliency}, etc. Mathematically, it is a simplicial complex structure to fit a 3D object, and maintains more important geometric features such as topology, curvature, and edges to support accurate geometric analysis \cite{surazhsky2005fast}\cite{lv2019nasal}. Comparing to a point cloud, a mesh provides more comprehensive representation. Although it has many advantages, a mesh is difficult to be achieved. Most of the 3D scanning devices can only provide point clouds, not meshes. Few devices support direct mesh acquisition \cite{kumar20183d} from real world and such devices are bulky and expensive. As a consequence, point cloud based 3D mesh reconstruction has emerged as one of the most fundamental techniques of 3D modelling.

In general, there are three key requirements involved in mesh reconstruction from point clouds, including local region detection, geometric feature keeping, and resampling. For the first requirement, according to the MLS (moving least squares) surface theory \cite{Alexa2001}, the surface of a 3D object can be divided as local regions, which are represented by neighbor relationship of points in a point cloud. However, such relationship is usually missing in raw data captured by scanning devices and thus needs to be restored for further processing using local region detection. For the second requirement, geometric features of original 3D objects should be kept in the reconstructed mesh, which are useful for geometric analysis \cite{surazhsky2005fast}\cite{lv2019nasal}. For the third requirement, resampling refers to reconstruction toward an isotropic 3D mesh with a given point number. In an isotropic mesh, the distances between each point to its neighbor points are approximately
equal in the mesh. With the same point number, isotropic meshes share a regular form which is useful for applications such as mesh-based shape registration \cite{bouaziz2016modern}, animation \cite{agudo2018shape}, and retrieval \cite{sahilliouglu2016detail}.

Most existing methods attempt to detect local regions in the local tangent space, such as Centroidal Voronoi Tessellation (CVT) \cite{Chen2018CVT} and Gaussian kernel \cite{zhong2019surface}. In the local tangent space, the neighbor relationship of points can be conveniently constructed and optimized. However, some geometric features such as external and internal edges could be lost. External edges means the sharp edges outside the 3D object and internal edges are inside the 3D object like holes. Some existing methods focus on certain geometric feature rebuilding in mesh reconstruction \cite{yadav2018robust}\cite{aroudj2017visibility}.
Focusing on certain geometric feature keeping in reconstruction, such methods have poor versatility and robustness generally. Most existing methods are incapable of reconstructing an isotropic mesh with the number of points specified by a user \cite{cohen2004greedy}\cite{Digne2011Scale}\cite{kazhdan2013screened}. It is a big challenge for a mesh reconstruction method to meet all the three aforementioned requirements. We propose a voxel structure-based mesh reconstruction to meet the three requirements.

The proposed two-step voxel structure based framework includes initial mesh reconstruction and mesh optimization. Firstly, the framework harvests an initial reconstructed mesh by the intrinsic metric with which means the distances between different points are defined by the geodesic distance instead of Euclidean distance \cite{Grossmann2002}. This improves the accuracy of local region detection and avoids the wrong neighbor relationship of points. Based on local regions, an initial mesh can be reconstructed. Secondly, the framework optimizes the initial reconstructed mesh into an isotropic one with geometric features. The point number does not change in the final reconstructed mesh. The key contributions of our work are summarized as follows.

\begin{figure*}
  \includegraphics[width=\linewidth]{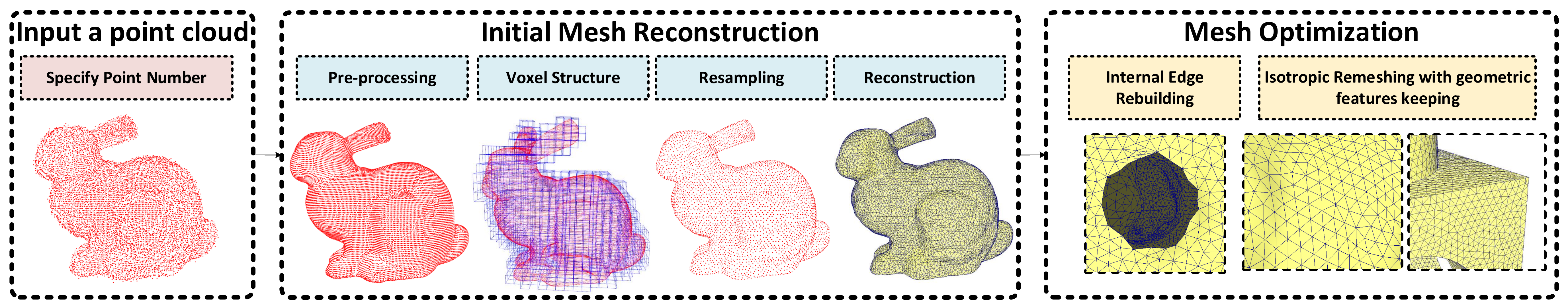}
  \caption{The pipeline of the proposed mesh reconstruction method.}
  \label{f1}
\end{figure*}

\begin{itemize}
\item We propose a voxel structure to reconstruct an initial mesh from a point cloud. It provides the intrinsic metric that improves the accuracy of the initial reconstructed mesh with any given point number.

\item Based on the voxel structure, we design a mesh optimization method for mesh quality improvement. With the proposed method, the initial reconstructed mesh can be optimized into an isotropic one while the important geometric features are kept.

\end{itemize}

The rest of the paper is organized as follows. In Sec. 2, we review existing classical methods for mesh reconstruction. We introduce the proposed initial mesh reconstruction method in Sec. 3, followed by the mesh optimization in Sec. 4. We demonstrate the effectiveness and efficiency of our method with extensive experimental evidence in Sec. 5, and Sec. 6 concludes the paper.

\section{Related Works}

There are different kinds of meshes such as triangular meshes \cite{shewchuk2002delaunay}, quadrilateral meshes \cite{blacker1991paving}, non-manifold meshes \cite{de2004multi}, and large-scale meshes \cite{lafarge2012creating}. In this paper, we focus on single 3D object triangular mesh reconstruction with the 2-manifold property. The related work can be broadly classified into three categories: Approximation-based, Delaunay-based, and Point resampling-based, as well as pre-processing and post-processing.

\subsection{Approximation-based Approaches}
Approximation-based methods attempt to rebuild the 2-manifold mesh to fit a point cloud directly. The methods achieve the reconstruction by establishing an objective function, such as Poisson Function~\cite{kazhdan2006poisson}\cite{kazhdan2013screened}, Scale Space \cite{Digne2011Scale}, MLS optimization \cite{Yutaka2005Sparse}\cite{Yukie2009Smoothing}, Implicit Function~\cite{Fuhrmann2014floating}, Voxel-based reconstruction~\cite{wang2005reconstructing}\cite{guan2020voxel}, and Kernel Density Estimation~\cite{aroudj2017visibility}. The Poisson function was a classical method for mesh reconstruction~\cite{kazhdan2006poisson}. The core idea was to obtain an indicator function by solving a Poisson formulation, which is a piece-wise constant function and signs the different sides of the surface. Scale space provided a smoothing operator for raw point clouds. By estimating the mean curvature and solving a mean curvature motion in point clouds, a smooth surface was constructed~\cite{Digne2011Scale}. The MLS-based methods \cite{Yutaka2005Sparse}\cite{Yukie2009Smoothing} reconstructed a 2-manifold mesh based on local surface fitting. In summary, such methods reconstruct a smooth surface from a point cloud and are robust to noise. However, the local shape features are broken to a certain degree and the stability is not satisfied caused by the wrong estimation for face normal and incorrect approximate region.

\subsection{Delaunay-based Approaches}
The Delaunay-based framework is regarded as the mainstream technology in 3D triangular meshing. It provides a simple and efficient point connect scheme for a point cloud without local surface approximation. A lot of mesh reconstruction methods based on Delaunay triangulation and their variations have been extensively investigated in the literature. Ameta~\cite{Amenta1988Voronoi} utilized the dual characteristics between Voronoi Diagram and Delaunay triangulation to rebuild the surface. He also provided some improvements, which are based on the crust algorithm to cover more geometric features in meshing~\cite{Amenta2000Simple} \cite{Amenta2001Power}.
Cohen~\cite{cohen2004greedy} proposed a greedy Delaunay triangulation to fix the local errors in the reconstructed mesh. Kuo~\cite{Kuo2005Delaunay} combined the advantages of Delaunay-based and region-growing methods to reconstruct the mesh. Peethambaran~\cite{Peethambaran2015reconstruction} built a graph structure to reconstruct a water-tight surface. Wang~\cite{Wang2019R} proposed a new triangle selection strategy to reduce the topology errors and holes in Delaunay triangulation. Liu \cite{liu2018shape} extended the Delaunay-based strategy to image stitching. Such methods are restricted by the points' positions in general. The quality of the reconstructed mesh is poor when the points' distribution is nonuniform in a point cloud.

\subsection{Point Resampling-based Approaches}
Considering the drawback of the Delaunay-based methods, some studies attempt to resample a point cloud into an isotropic one. As discussed in being isotropic means that the distances between each point to its neighbor points are approximately equal. For a point cloud, such distances from all points form a distance field. The resampling process is used to optimize the distance field to achieve an isotropic mesh. The representative method was Centroidal Voronoi Tessellation (CVT)\cite{Liu2009C}\cite{Chen2018CVT} based resampling. Based on the Lloyd's relaxation, the Voronoi Diagram was optimized in local tangent space~\cite{Yan2009CVT}. The iteration of Lloyd relaxation required heavy calculation to update the Voronoi Diagram and new centers. Some researchers \cite{FEI2014P}\cite{BOLTCHEVA2017123}\cite{ray2018meshless} attempted to improve the effectiveness based on parallel computation. Some other methods used different kinds of point sampling schemes to achieve a similar result. Luo~\cite{Luo2018U} proposed a point cloud resampling method based on the Gaussian-weight Laplacian graph. Zhong~\cite{zhong2019surface} used a Gaussian kernel function to achieve the isotropic point cloud for mesh reconstruction. Such methods optimize the distance field of the point cloud before meshing. The advantages of the methods include the high quality of the triangulation results and the robustness to different local points' density. However, some local geometric features are lost during optimization. In our framework, the same scheme is adopted to achieve an initial reconstructed mesh from a point cloud resampling result. The distance field of the resampling result is optimized in the voxel structure. Our reconstruction method is classified into this part and attempt to keep geometric features in the reconstructed mesh.

\subsection{Pre- and Post-processing}

The approaches for pre-processing and post-processing included mesh denoising \cite{fan2009robust}\cite{wang2016mesh}\cite{yadav2018robust}\cite{wang2019data}, isotropic remesh \cite{botsch2004remeshing}\cite{levy2010p}\cite{2013Adaptive}\cite{wang2018isotropic}\cite{du2018field} and mesh repair \cite{ju2004robust}\cite{hetroy2011mesh}\cite{zhou2016mesh}\cite{chu2019repairing}. Such methods are used to improve the mesh quality based on geometric evaluations \cite{lavoue2010comparison}. In our framework, we provide similar functions in mesh
optimization. Using mesh optimization with the voxel structure, the quality of a reconstructed mesh can be improved.

\section{Initial Mesh Reconstruction}

As mentioned in Sec. 2, the initial mesh reconstruction task in our framework can be formulated as a point-based distance field optimization problem. Once the distance field is optimized, the local regions are convenient to be detected and the initial reconstructed mesh can be obtained. We propose a voxel structure-based framework to optimize the distance field. The framework mainly includes three main components: point cloud pre-processing, construction of voxel structure, and initial mesh reconstruction. The technical details of each component will be elaborated in the following subsections.

\begin{figure}
  \includegraphics[width=\linewidth]{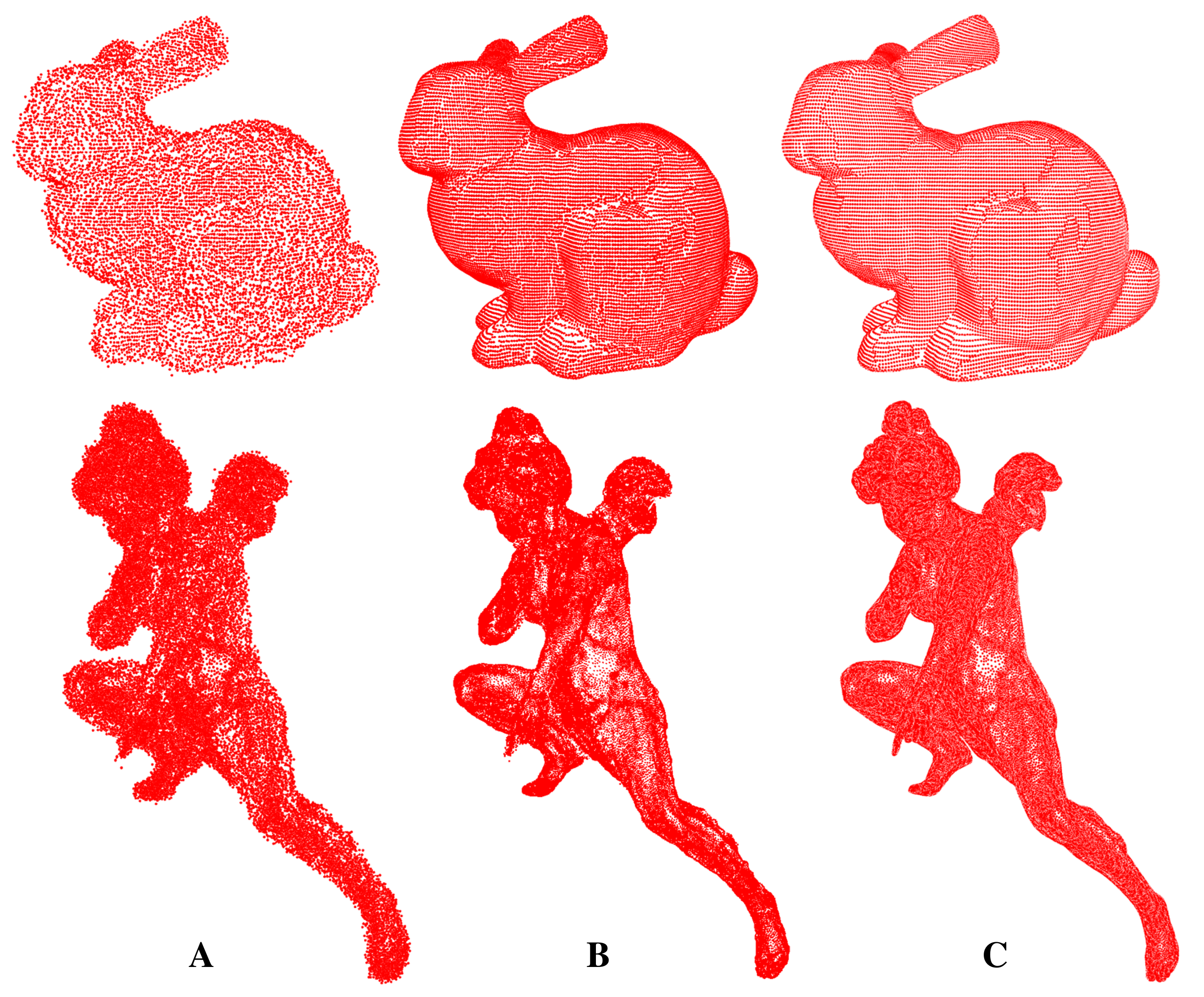}
  \caption{Instances of denoising results. A: Point cloud models with noise; B: Denoising results; C: Ground truth.}
  \label{f2}
\end{figure}

\subsection{Point Cloud Pre-processing}

Limited by the accuracy of 3D scanning devices, raw point clouds generally carry noise and are with uncertain point density distribution, which could affect the points-based distance field optimization. Therefore, a point cloud pre-processing is needed for input point cloud denoising and uniform density adjusting. For point cloud denoising, we use an MLS surface-based smoothing method \cite{Alexa2001}. It is formulated in Equation \ref{en1}. The parameters $p_i$ and $p_j$ represent the points from a point cloud $P$, $p_j$ is the neighbor point of $p_i$ and $p_{i}'$ is the new position of point $p_i$ after least squares fitting. It is achieved by weighted fitting from its neighbors. The weight function $\theta$ in Equation \ref{en2} is suggested to fit a Gaussian function \cite{levin2004mesh}. The $h$ is a fixed parameter reflecting the anticipated spacing between neighboring points. In our paper, we compute distances $\{d_k(p_i)\}$ between all points and their $k$-th neighbors , $h = max\{d_k(p_i)\}$. Such processing can be regarded as a "pulling back" mapping from points to the MLS surface. In Figure \ref{f2}, we show two denoising instances.

\begin{equation}
\label{en1}
p_{i}' ={\textstyle\sum_{j=0}^{k-1}}\frac{\theta(\left|\left|p_i-p_j\right|\right|)p_j}{\left|\left|\sum_{j=0}^{k-1}\theta(\left|\left|p_i-p_j\right|\right|)\right|\right|}
\end{equation}

\begin{equation}
\label{en2}
\theta(d)=e^{-\frac{d^2}{h^2}}
\end{equation}

\begin{figure}
  \includegraphics[width=\linewidth]{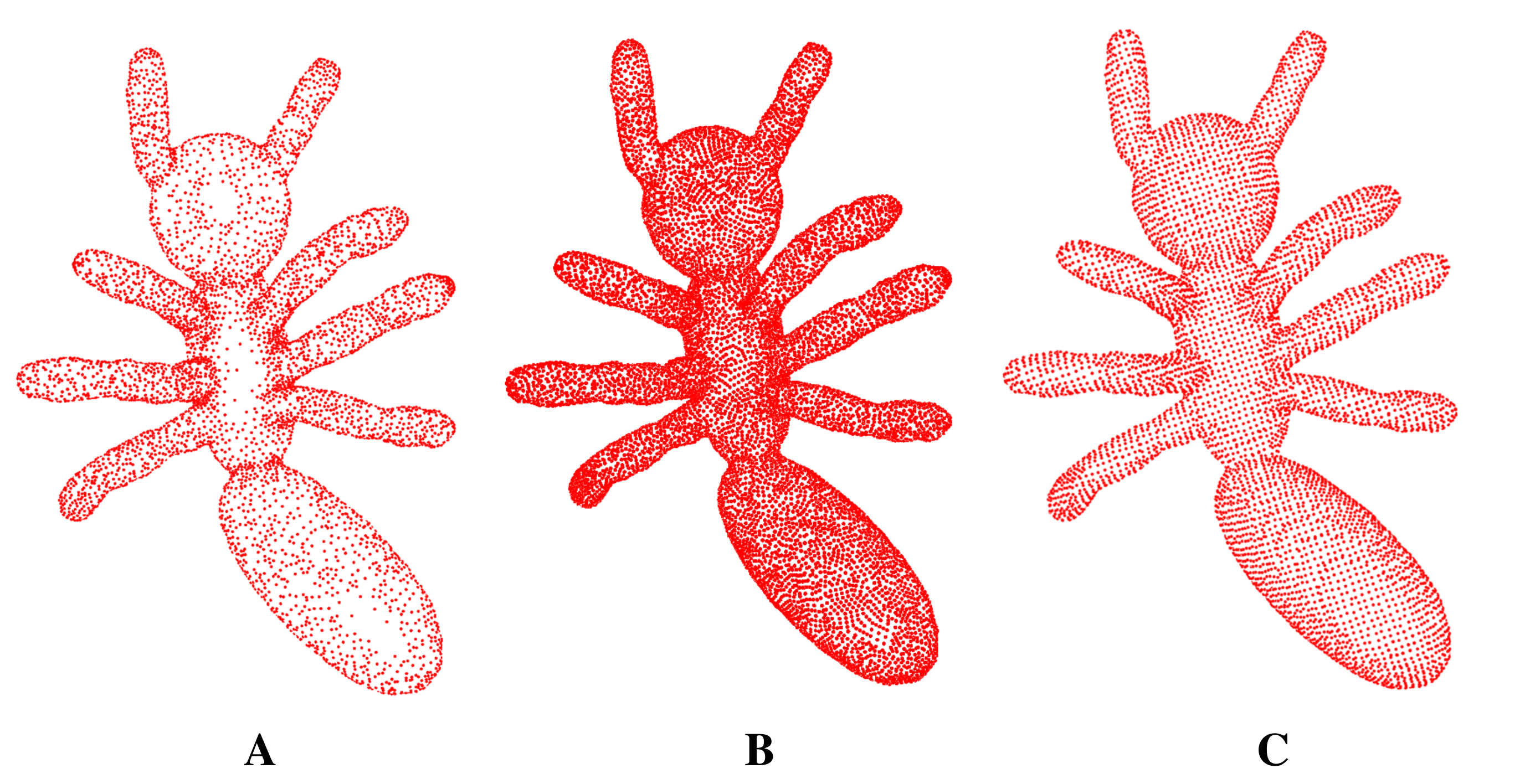}
  \caption{An instance of octree-based density adjusting. A: A sparse point cloud with non-uniform density; B: Delaunay-based interpolation for point cloud up-sampling; C: octree-based density adjusting result.}
  \label{f3}
\end{figure}

For uniform density adjusting, we utilize an octree-based method \cite{schnabel2006octree}. It is used to adjust the density of a point cloud based on the voxel boxes. The scale of the voxel box controls the distances between different points, which uniforms the points' density in a point cloud. In Figure \ref{f3}, we show an instance of octree-based density adjusting. We compute the distance set $\{d_n(p_i)\}$ between each point to its nearest neighbor point. The scale is selected by the mean value from $\{d_n(p_i)\}$. In Figure \ref{f3}C, an instance of uniform density adjusting is shown. The local region can be detected from the uniform density point cloud conveniently in following step.

\begin{figure}
  \includegraphics[width=\linewidth]{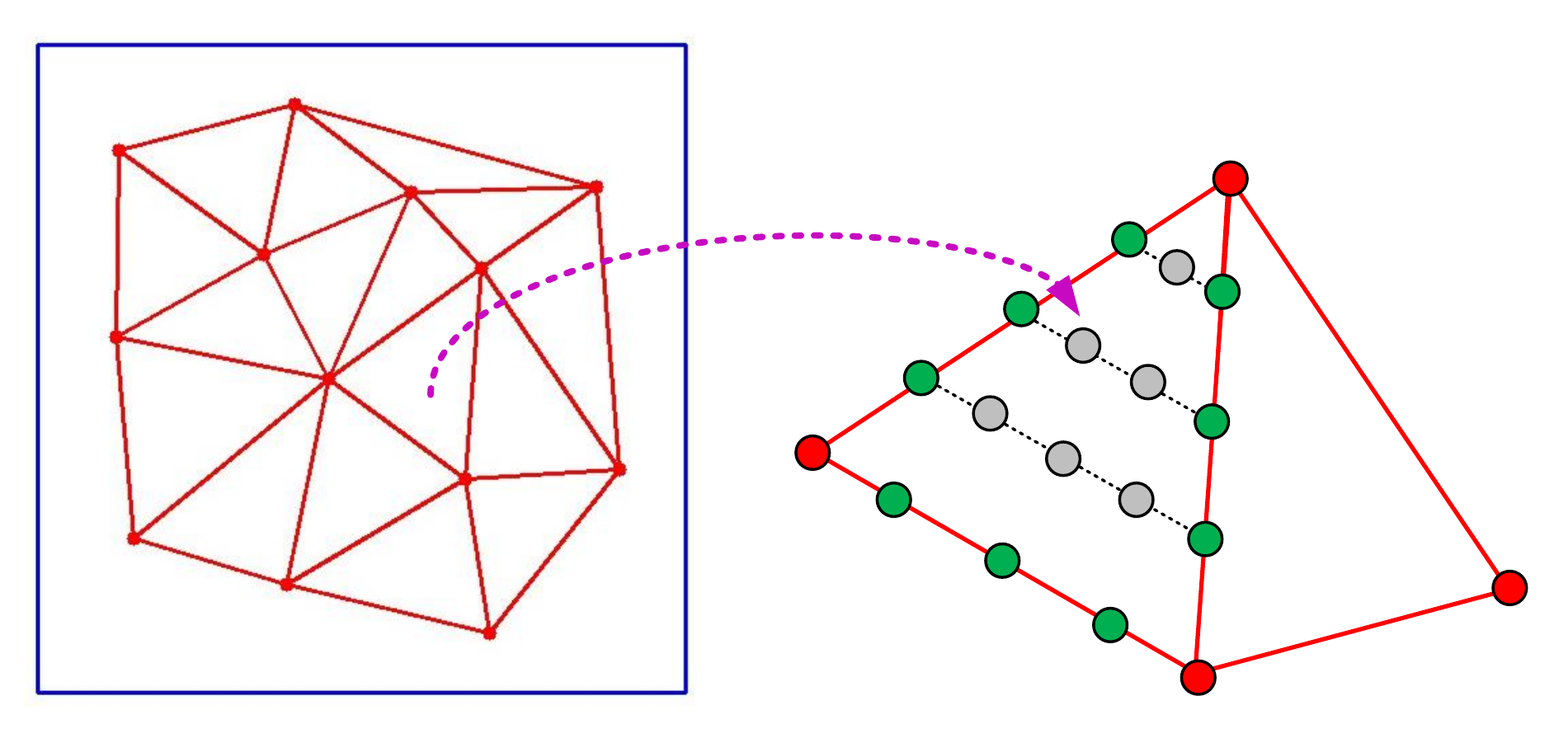}
  \caption{An instance of the up-sampling. The gray points represent new points ($s = 6$) by the up-sampling. The points are proportional to the triangle's area. The green points in single edge ($s_l = 3$) are computed by $s$.}
  \label{f4}
\end{figure}

Using the octree-based density adjusting, the point number of a point cloud will be reduced. In some cases (sparse point clouds or user-specified point number is similar to a given point cloud), we would not like to reduce the point number while adjusting the point cloud density. Then an up-sampling is needed before the density adjusting. We propose an up-sampling processing based on Delaunay-based interpolation. The Delaunay-based interpolation builds the Delaunay triangulation region for each point and inserts the new points into the region. The up-sampling points are inserted into the triangle's edges at first. The insert point number in single edge $s_l$ as defined in Equation \ref{en3} (s = 6 in default). Then we insert the points into the triangle region as shown in Figure \ref{f4}. In Figure \ref{f3}B, we show the up-sampling instance. After the pre-processing, a point cloud can be used for the next steps.

\begin{equation}
\label{en3}
s_l=\left[\sqrt{2s+1/2}-1/\sqrt2\right]
\end{equation}

\subsection{Construction of Voxel Structure}

\begin{figure}
  \includegraphics[width=\linewidth]{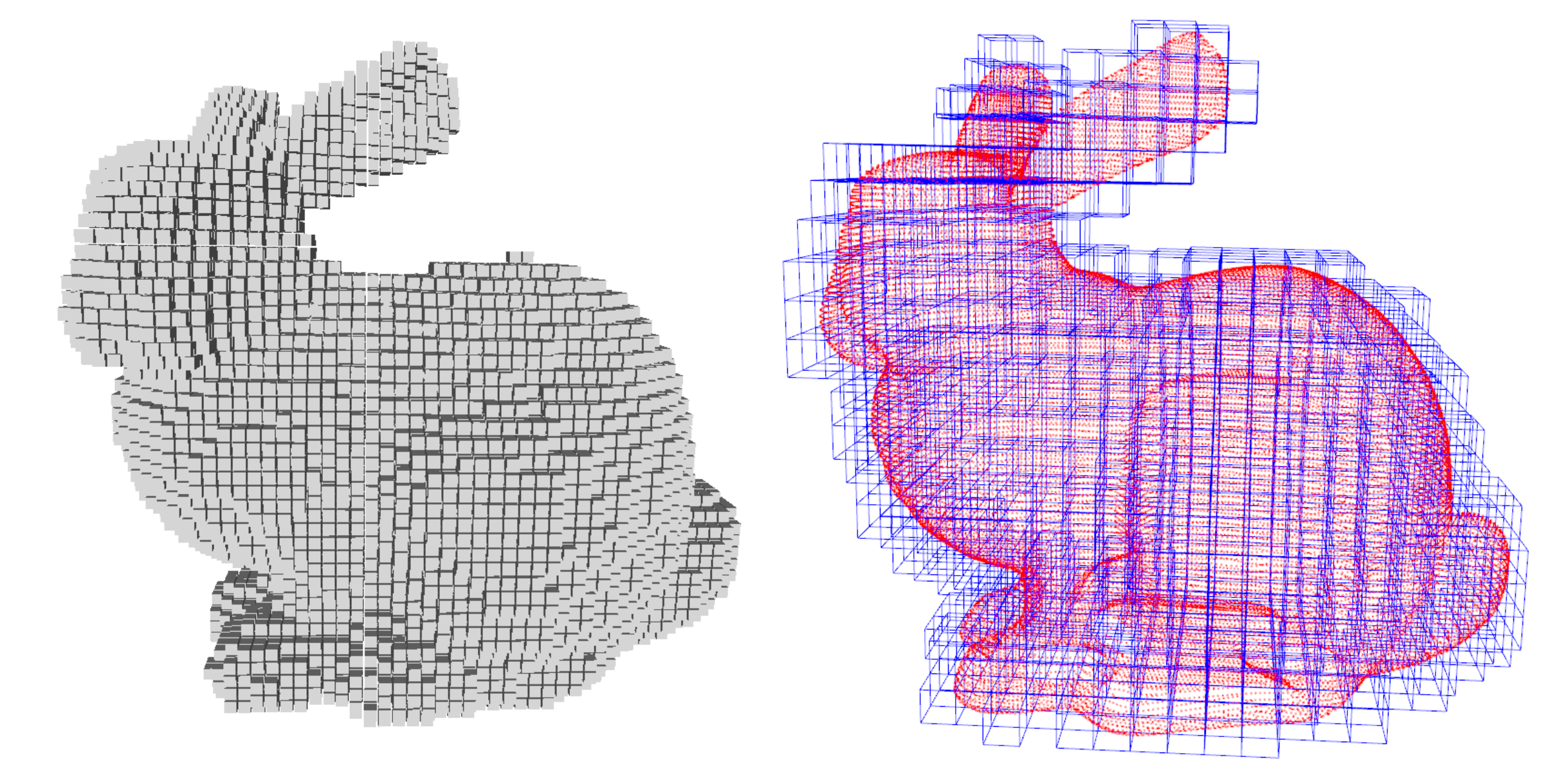}
  \caption{Comparison between voxelization and voxel structure}
  \label{f5}
\end{figure}

\begin{figure}
  \includegraphics[width=\linewidth]{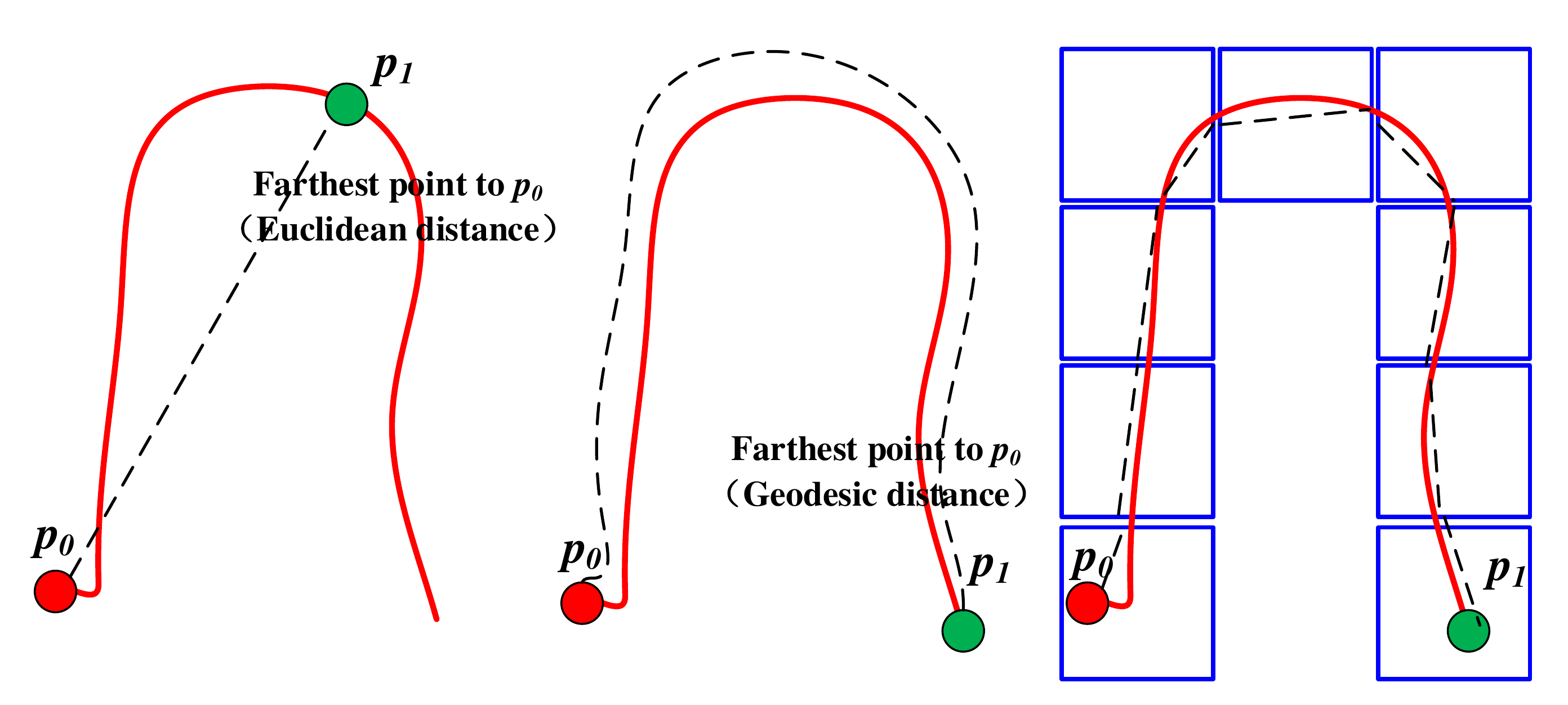}
  \caption{Comparison of Euclidean distance and geodesic distance. The geodesic distance reflects the real distance between $p_0$ and $p_1$ on the curve. Based on the voxel structure, the global geodesic distance is fitted by the sum of local Euclidean distance in each voxel box.}
  \label{f6}
\end{figure}

The voxel structure is proposed for the point-based distance field optimization. It is similar to the octree and voxelization. In Figure \ref{f5}, we compare the voxelization and the voxel structure. The points are aggregated into voxel blocks by voxelization, but divided into different voxel boxes in the voxel structure. The voxel structure provides an intrinsic metric for a point cloud to avoid the incorrect local region detection by Euclidean distance. In Figure \ref{f6}, we compare the two different distance definitions.

The intrinsic metric is based on the geodesic distance, which is similar to the shortest path search between different voxel boxes in the voxel structure \cite{Grossmann2002}. Inside a voxel box with a small scale, curvature transfer of points is controlled in a small range. The geodesic distance is similar to the Euclidean distance, and therefore approximate errors are decided by the scale of the voxel box. Once the scale is determined, the voxel structure is established. The scale should balance the approximate error and point number in different boxes (if the scale is too small, there will be no enough points in a single voxel box for the further process). We provide a scale calculation formula as a default suggestion. In Equation \ref{en4}, we show the formula of the scale $v_{scale}$. The length $l$ is the longest border of the bounding box from a point cloud $P$ with $\vert P\vert$ points.

\begin{equation}
\label{en4}
v_{scale} = \left[\frac{2\times l}{\sqrt[3]{\vert P\vert}}\right]
\end{equation}

\subsection{Initial Mesh Reconstruction}

Based on the voxel structure, we provide a point cloud resampling scheme for initial mesh reconstruction. The resampling optimizes the distance field. The point number after resampling is equal to the user-specified number. Based on the resampling result, the initial reconstructed mesh can be obtained.

In our scheme, the resampling is processed in different voxel boxes independently. The geodesic computation can be approximated by Euclidean measure. It reduces the computation cost while keeping the intrinsic metric for resampling. Following the property, we design the resampling scheme to be a sub-division strategy. The resampling task for a point cloud is divided into local resampling tasks in different voxel boxes. According to the resampling rate $R_s$ or user-specified number $\left|P_s\right|$, the local resampling point number $\left|P_{vs}\right|$ in the voxel box $v$ is computed by proportional calculation in Equation \ref{en5}. The parameter $\left|P_v\right|$ represents the point number in $v$.

\begin{equation}
\label{en5}
\begin{array}{c}
\left|P_{vs}\right|=\left|P_v\right|\times R_s, \left|P_{vs}\right|\in N^+\\
R_s=\left|P\right|/\left|P_s\right|
\end{array}
\end{equation}

\begin{figure}
  \includegraphics[width=\linewidth]{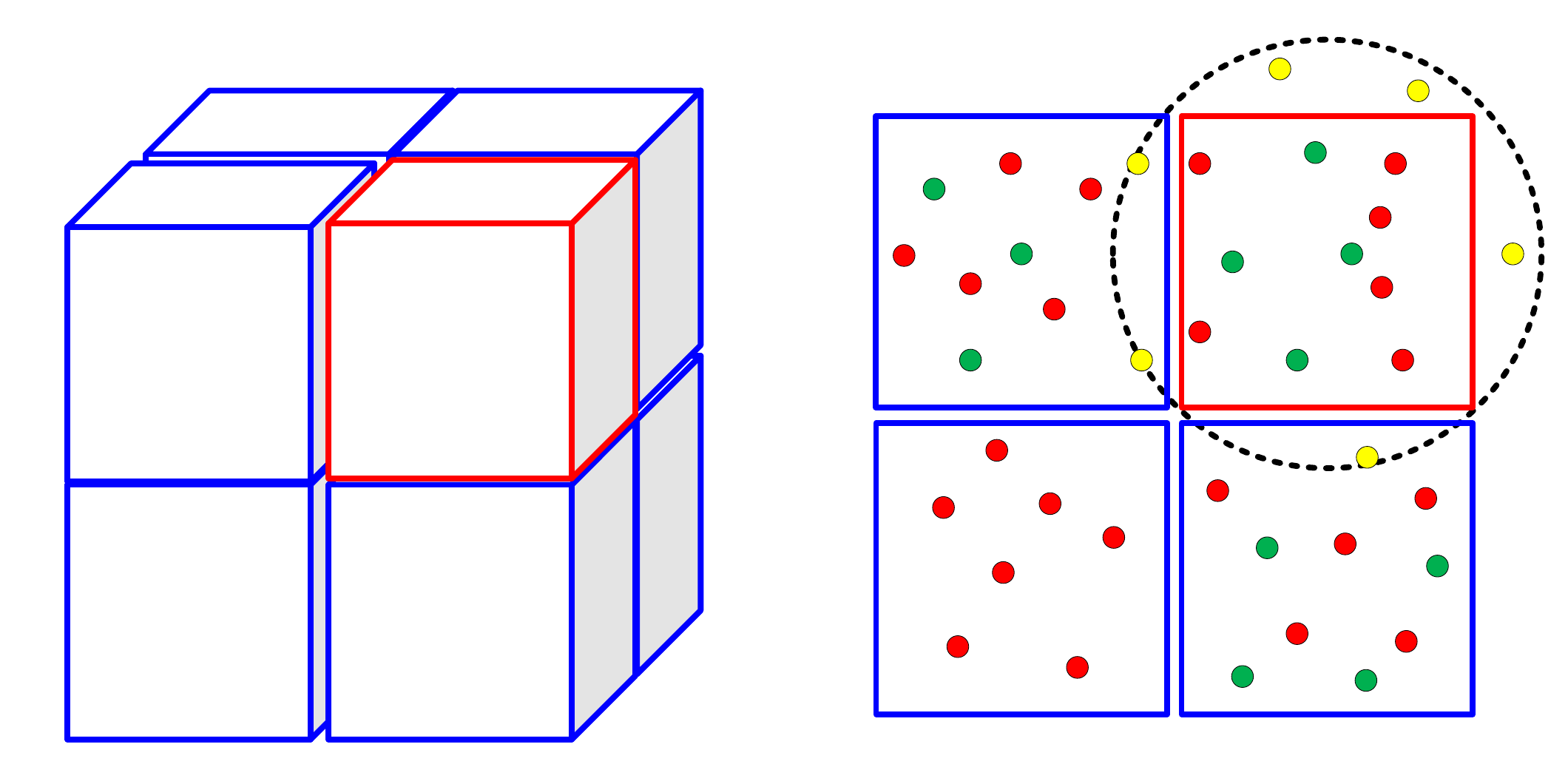}
  \centering
  \caption{An instance of the resampling scheme for a voxel box (red). The red points represent the original points in the point cloud, while the green ones are resampled points by FPS. The yellow points are resampled points in neighbor voxel boxes, which are considered by FPS for the red voxel box.}
  \label{f7}
\end{figure}

\begin{figure}
  \includegraphics[width=\linewidth]{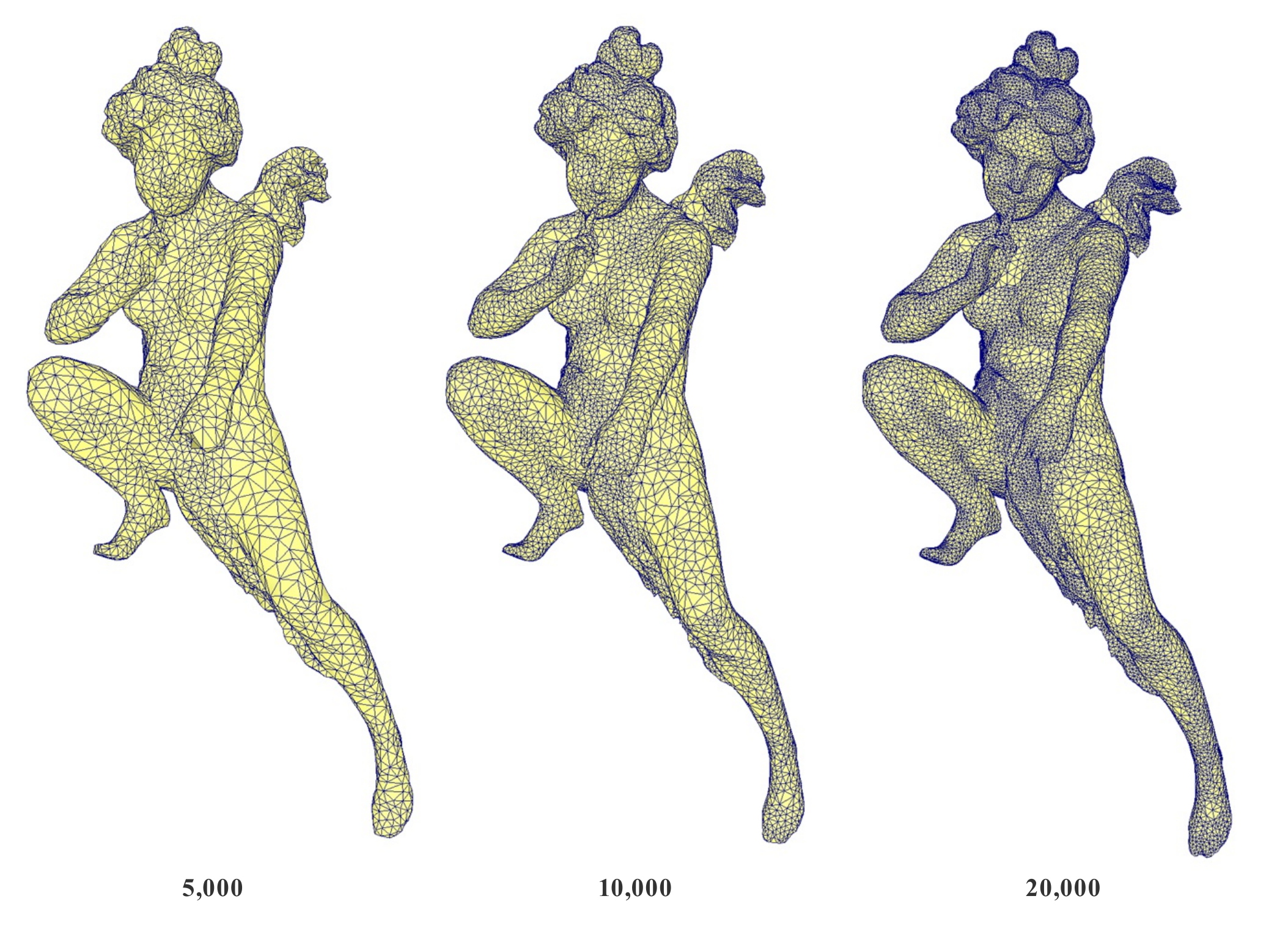}
  \centering
  \caption{An instance of the initial reconstructed mesh with different point numbers by user specification.}
  \label{f8}
\end{figure}

Once the local resampling point numbers for different voxel boxes are determined, the resampling can be processed in parallel. We use the Farthest Point Sampling (FPS) ~\cite{Moenning2003fast} to resample the point cloud in different voxel boxes. Using FPS can achieve an approximate isotropic local resampling result. Even the local resampling results satisfy the requirements of resampling, the final resampling result can not be achieved by the combination of local resampling results. The reason is that the distances between points in adjacent voxel boxes are not approximately the same. To solve the problem, we divide the parallel computation for local resampling into different rounds. The adjacent voxel boxes are not resampled in the same round. Besides, resampling points of neighbor boxes should be included by FPS for the processing voxel box. In Figure \ref{f7}, we show an instance to explain the resampling scheme. The parallel computation is divided into 8 rounds which guarantees the approximate isotropic property of resampling. Based on the resampling result, we use a Delaunay triangulation method \cite{cohen2004greedy} to reconstruct mesh. Since the distance field is optimized with intrinsic metric, the reconstructed mesh is more accurate. In Figure \ref{f8}, we show a reconstruction instance with different point numbers by user specification.

\begin{figure}
  \includegraphics[width=\linewidth]{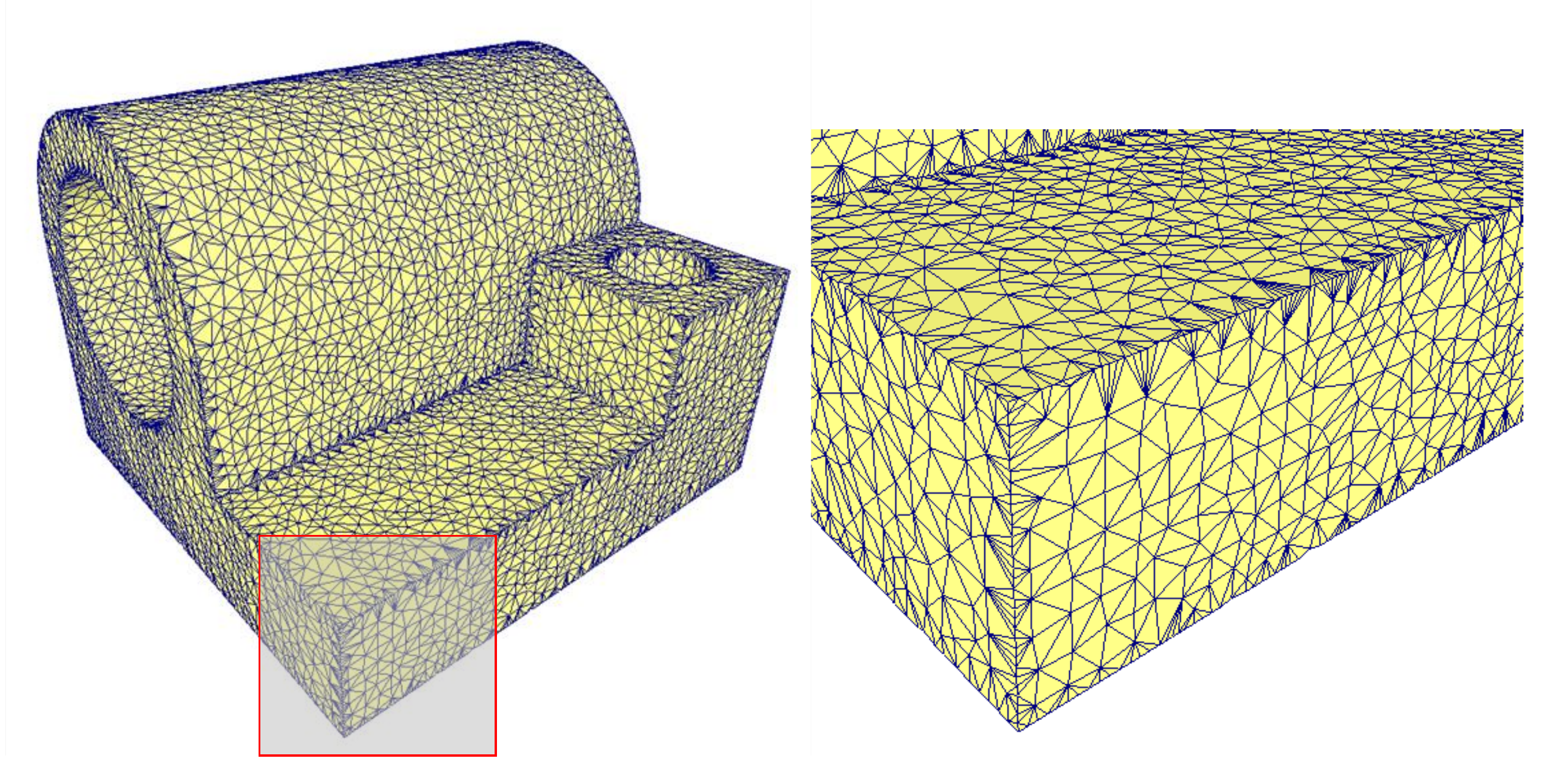}
  \centering
  \caption{An instance of reconstructed mesh with external edges.}
  \label{f9}
\end{figure}

\begin{figure}
  \includegraphics[width=\linewidth]{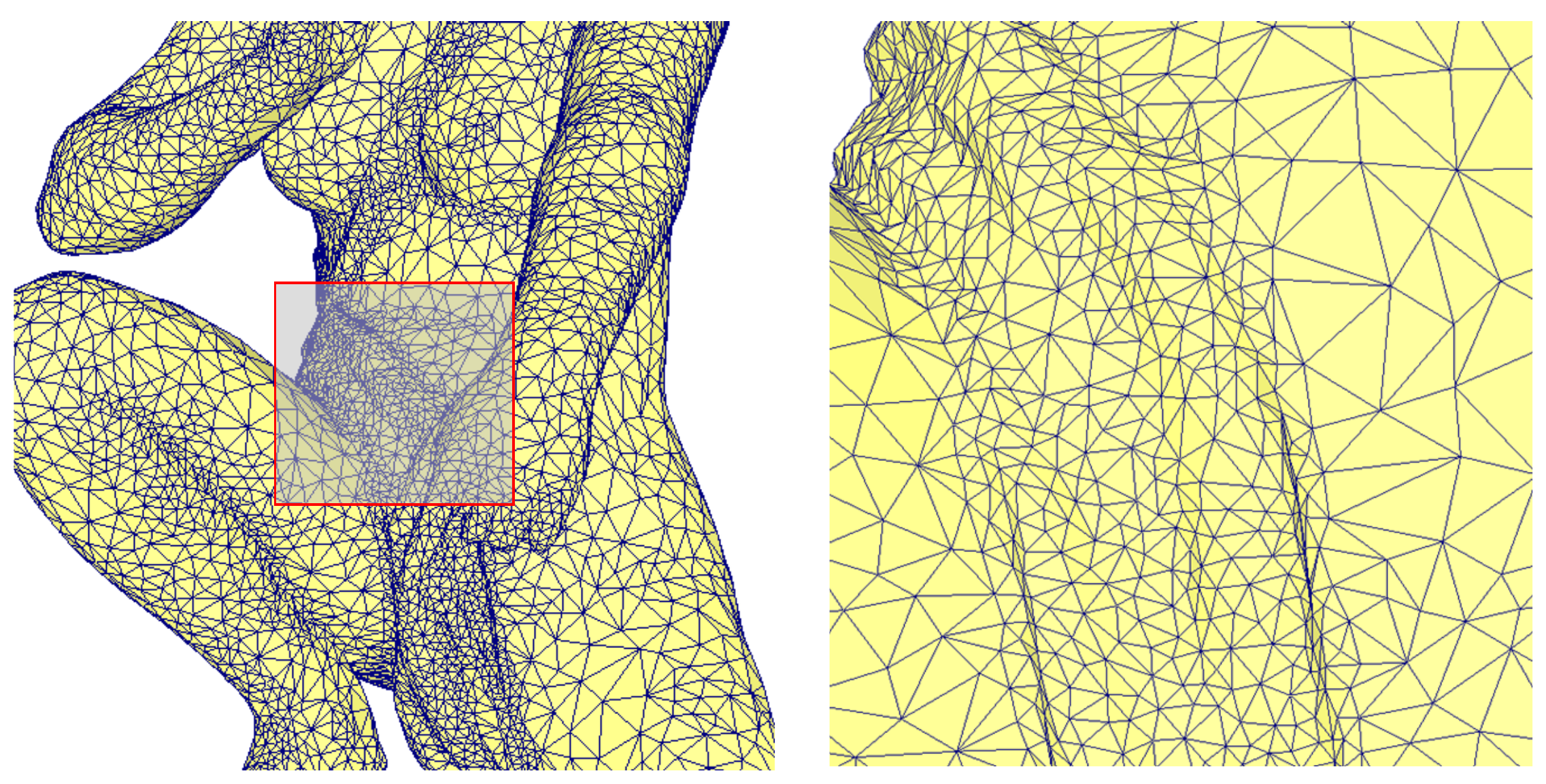}
  \centering
  \caption{An instance of reconstructed mesh with curvature sensitive keeping. The points are classified into five categories based on curvature values, the  proportion of the rates is 2:3:4:5:6.}
  \label{f10}
\end{figure}

To keep important geometric features, the resampling scheme provides a convenient solution by specifying different resampling rates for different point sets. Based on the different resampling rates, the target geometric feature can be maintained in the reconstructed mesh. For instance, if we want to keep external edges, the resampling rate for the points on the external edge should be larger than other points. It guarantees the quality of external edges in the initial reconstructed mesh. The points are classified into two categories (edge points and ordinary points) with two user-specified resampling rates. The edge points can be detected by \cite{merigot2010voronoi}. According to the rates, the local resampling is processed twice and external edges are kept. In practice, we set the proportion of two rates to 7:3. In Figure \ref{f9}, we show an instance of the mesh reconstruction with external edge keeping. For curvature sensitive keeping, it can be processed by the same scheme. The points are classified by curvature value. Using different resampling rates, the curvature sensitive initial reconstructed mesh can be achieved. In Figure \ref{f10}, we show an instance of the mesh reconstruction with curvature sensitive keeping.

\section{Mesh Optimization}

After the processing in the voxel structure, an initialized reconstructed mesh is achieved from the point cloud. The intrinsic metric of the voxel structure avoids the wrong adjacency in the initial reconstructed mesh. Geometric features such as external edges can be kept in the mesh if needed. To further improve the mesh quality, we provide a mesh optimization framework. The framework includes internal edge rebuilding and isotropic remeshing.

\subsection{Internal Edge Rebuilding}

\begin{figure}
  \includegraphics[width=\linewidth]{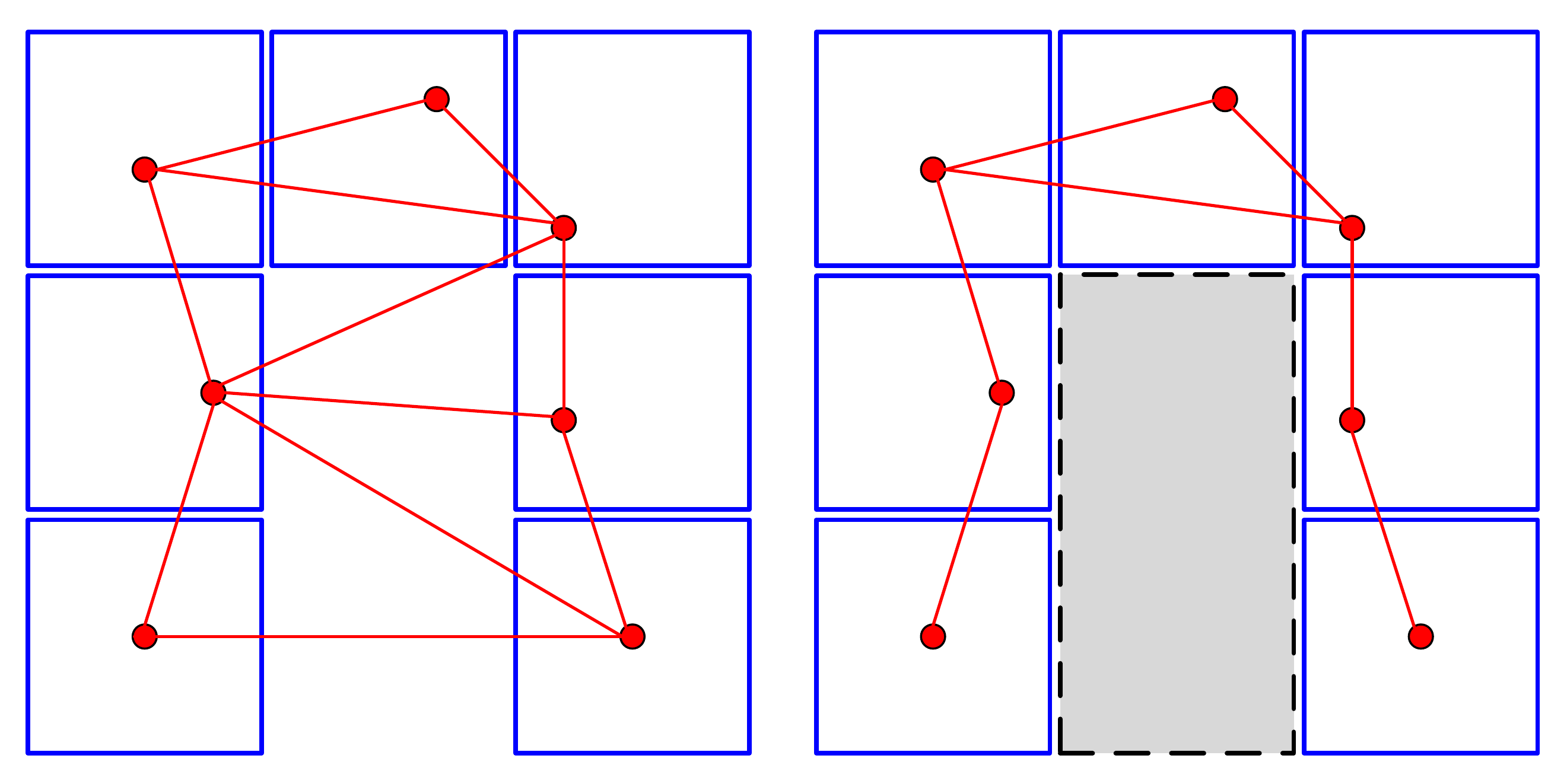}
  \centering
  \caption{An instance of illegal triangles which cross the non-adjacent voxel boxes.}
  \label{f11}
\end{figure}

\begin{figure}
  \includegraphics[width=\linewidth]{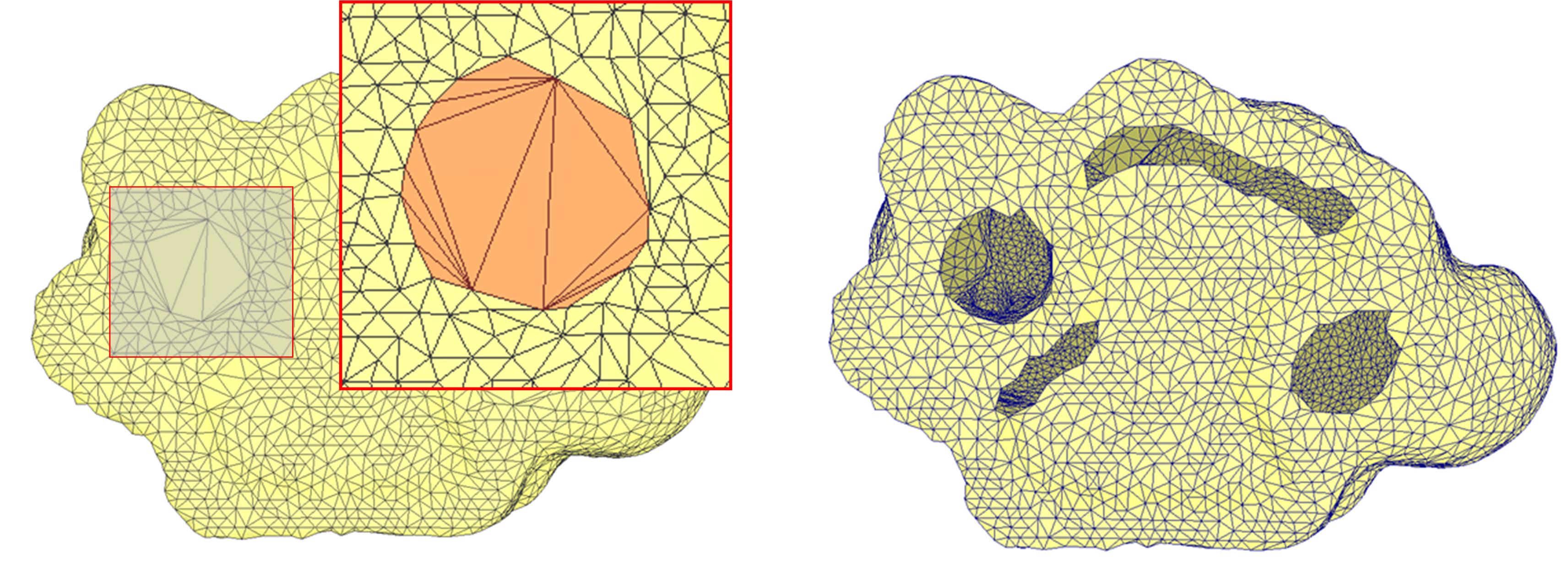}
  \centering
  \caption{An instance of internal edge rebuilding.}
  \label{f12}
\end{figure}

Internal edge is one kind of important geometric features. Compared to the external edge, the internal edge is inside a mesh like a hole, which affects the mesh topology structure. Most reconstruction methods fill the internal edge by default. In our framework, an internal edge rebuilding is provided as an optional function. The difficulty of the internal edge rebuilding is that the non-uniform density distribution of a point cloud affects the internal edge detection. After the density adjusting in the voxel structure, the point cloud is transferred into a uniform density one. The initial reconstructed mesh inherits the uniform density from the voxel structure. It means the triangle of the mesh should not cross the non-adjacent voxel boxes. In Figure \ref{f11}, we show the illegal triangles. Once such triangles are deleted from the mesh, the internal edges are rebuilt. In Figure \ref{f12}, we show an instance of internal edge rebuilding.

\subsection{Isotropic Remeshing}

Isotropic remeshing is used to optimize a mesh into an isotropic one. It is constructed by three basic steps: split, collapse, and flip \cite{botsch2004remeshing}, which are used to cut the long edge, delete the short edge, and optimize the valence of the mesh after split and collapse, respectively. Compared to the original isotropic remeshing, we change the implementations to achieve better results. The important changes include point number, collapse, and flip control. In the original isotropic remeshing, the point number can not be controlled precisely. We reduce the collapse frequency in each iteration to control the point number of the mesh. For external edge keeping, some collapse and flip should not be processed to avoid the feature lost. The remeshing includes the following steps:
\begin{enumerate}

\item Input initial reconstructed mesh with iteration number and point classification (external edge points and ordinary points).

\item Compute the average triangle edge length $l$ of the mesh.

\item Split the triangle edge if its length is larger than $\frac43 l$, count the split number $i_s$.

\item Collapse the triangle edge into the middle point if its length is smaller than $\frac45 l$. If the triangle edge includes an external edge point with more than two external edge neighbors, the collapse should not be processed. If the collapse number is equal to $i_s$, then stop.

\item Flip triangle edge to optimize the valence. If the flip generates obtuse angles or the flip triangle edge includes two external edge points, the flip should not be processed.

\item If the point is not the external edge point, smooth points in tangent space.

\item If the current iteration number is equal to the input one, then output the mesh. If not, loop to Step 2.

\end{enumerate}

\begin{figure}
  \includegraphics[width=\linewidth]{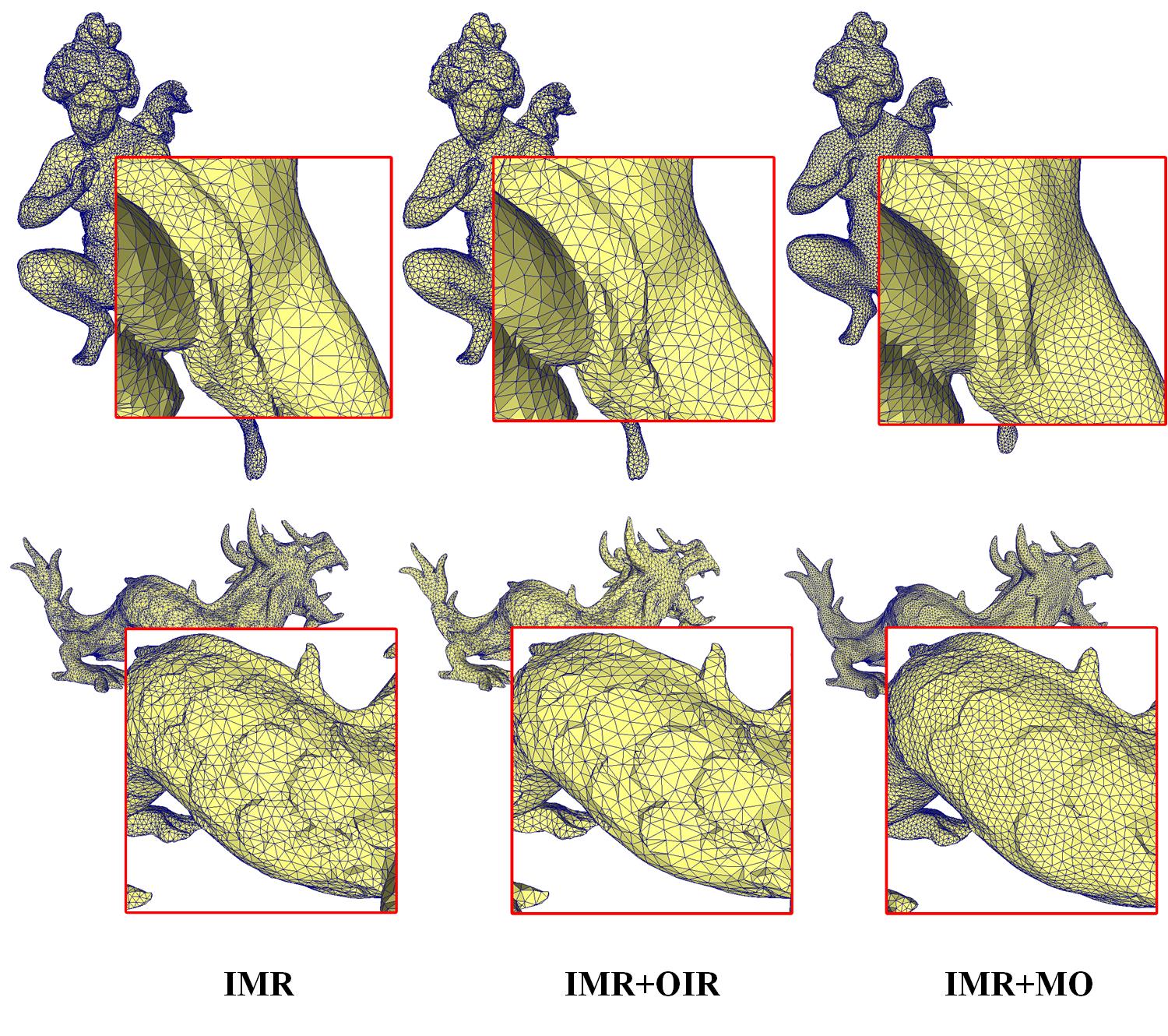}
  \centering
  \caption{Comparisons of mesh reconstruction results by IMR(initial mesh reconstruction), IMR+OIR(initial mesh reconstruction with original isotropic remeshing\cite{botsch2004remeshing}), and IMR+MO(initial mesh reconstruction with mesh optimization).}
  \label{f12_1}
\end{figure}

\begin{figure}
  \includegraphics[width=\linewidth]{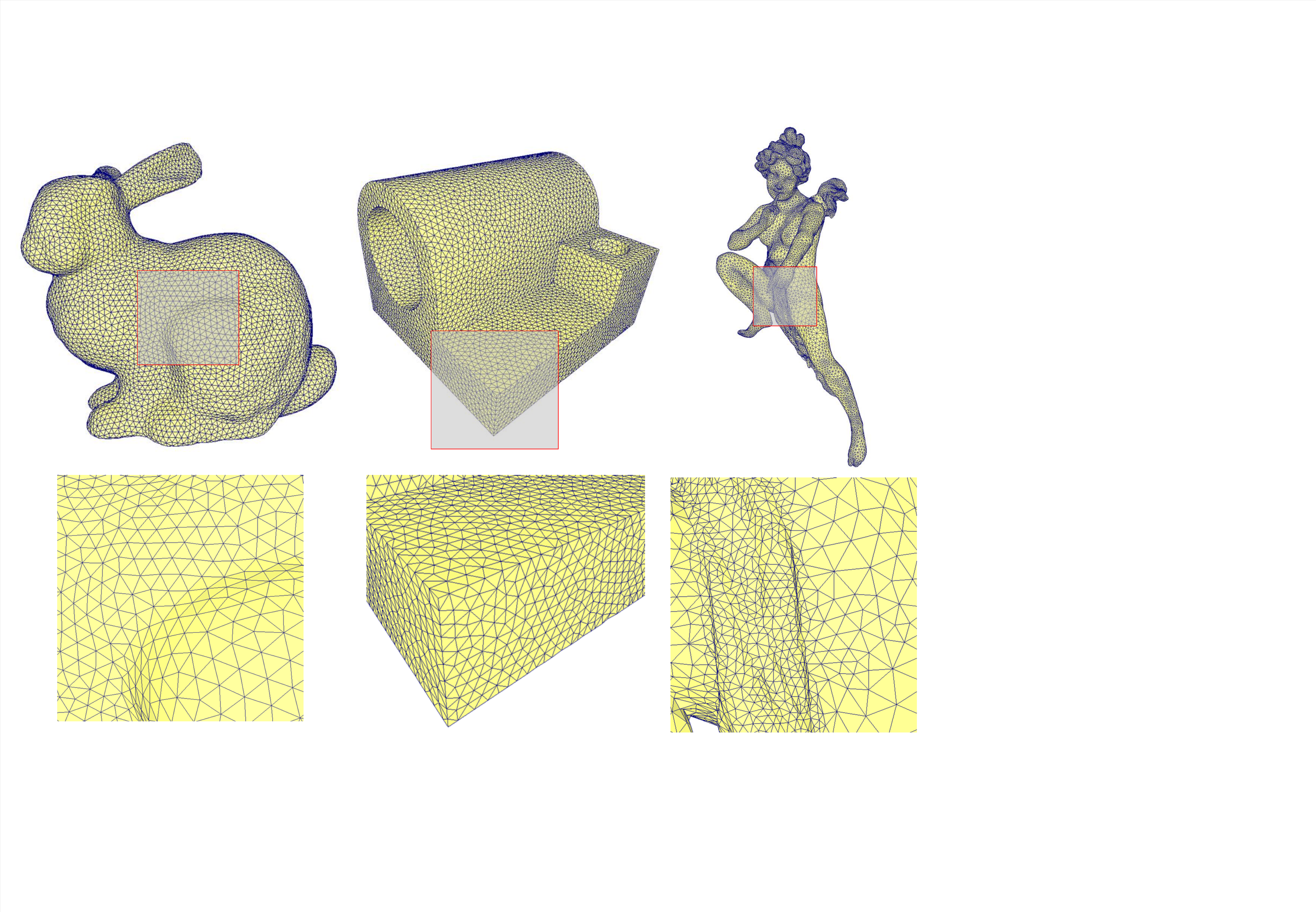}
  \centering
  \caption{Instances of isotropic remeshing. From left to right: isotropic remesh result; isotropic remesh result with external edges; adaptive isotropic result with curvature sensitive property.}
  \label{f13}
\end{figure}

The details of valence optimization and tangent space smoothing are introduced in \cite{botsch2004remeshing}. Based on the modification, the point number can be controlled and the generation of obtuse triangle is limited which improves the quality of the mesh.
Compared with the original isotropic remeshing method\cite{botsch2004remeshing}, our mesh optimization improves the quality of meshes. In Figure \ref{f12_1}, we compare the reconstructed meshes from initial mesh reconstruction (first step of our framework), initial mesh reconstruction with original isotropic remeshing\cite{botsch2004remeshing}, and initial mesh reconstruction with mesh optimization (our whole framework). It is clear that our framework achieves better isotropic property. For curvature sensitive keeping, the adaptive isotropic remeshing \cite{2013Adaptive} can be used to meet the requirement. The principle of adaptive isotropic remeshing is to change the length $l$ for different triangles. The details can be found in \cite{2013Adaptive}. In Figure \ref{f13}, we show
some isotropic remeshing instances. In summary, the mesh optimization improves the quality of the reconstructed mesh while keeping the important geometric features. Combining the initial mesh reconstruction and mesh optimization, our framework provides complete functions to solve the three key requirements of mesh reconstruction.

\begin{figure}
  \includegraphics[width=\linewidth]{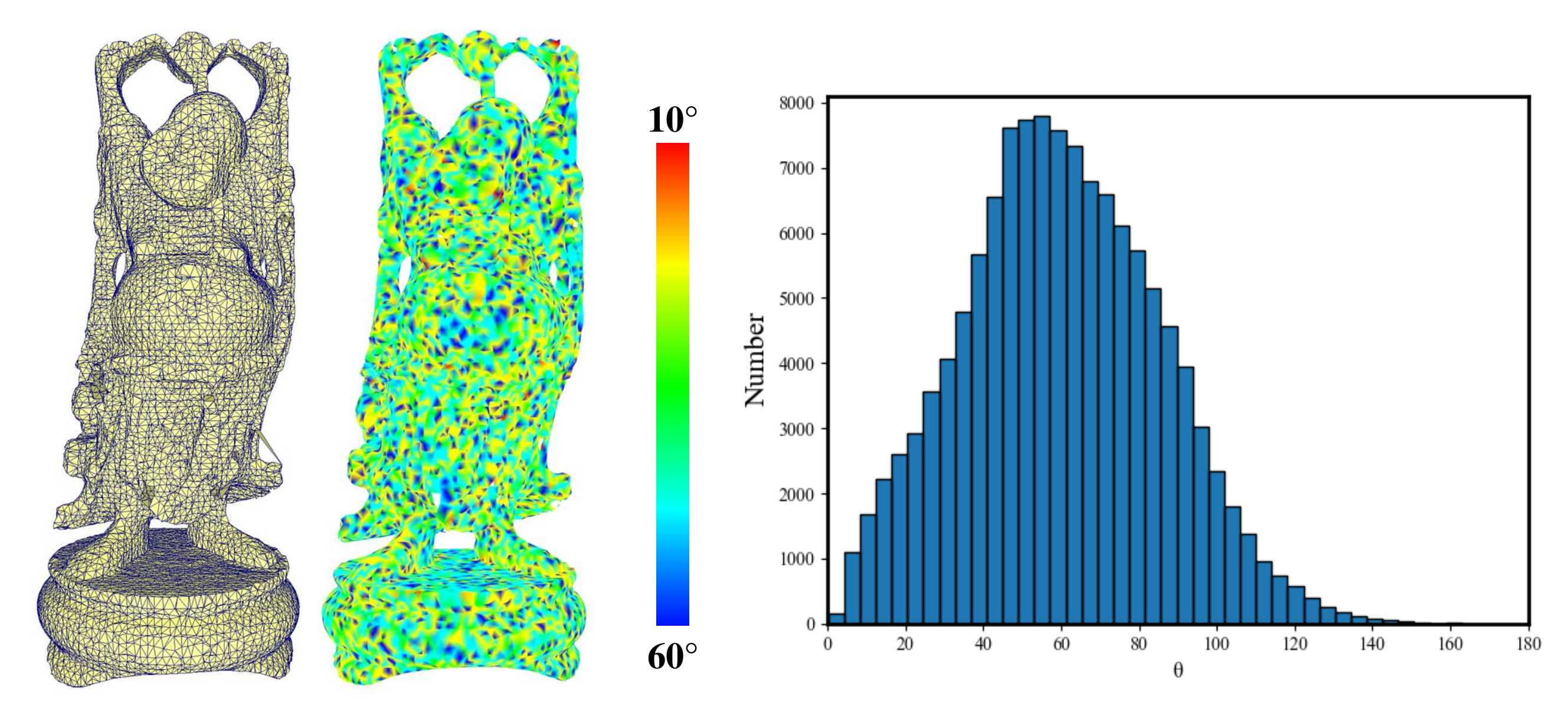}
  \caption{An instance of the color map and histogram for a reconstructed mesh (Buddha). }
  \label{f14}
\end{figure}

\section{Experiments}

We show the performance of our mesh reconstruction framework in this section. The experimental point cloud models were selected from Stanford and Shrec Models. We conducted the experiments on a machine equipped with Intel Xeon W 2133 3.6G Hz, 32 GB RAM, Quadro P620, and with Windows 10 as its running system and Visual Studio 2019 (64 bit) as the development platform. Firstly, we introduce the evaluation metrics for mesh quality measurement. Secondly, we compare several classic mesh reconstruction methods based on the evaluation metrics. Finally, we show a comprehensive analysis of our method with the state of art.

\begin{figure*}
  \center
  \includegraphics[width=\linewidth]{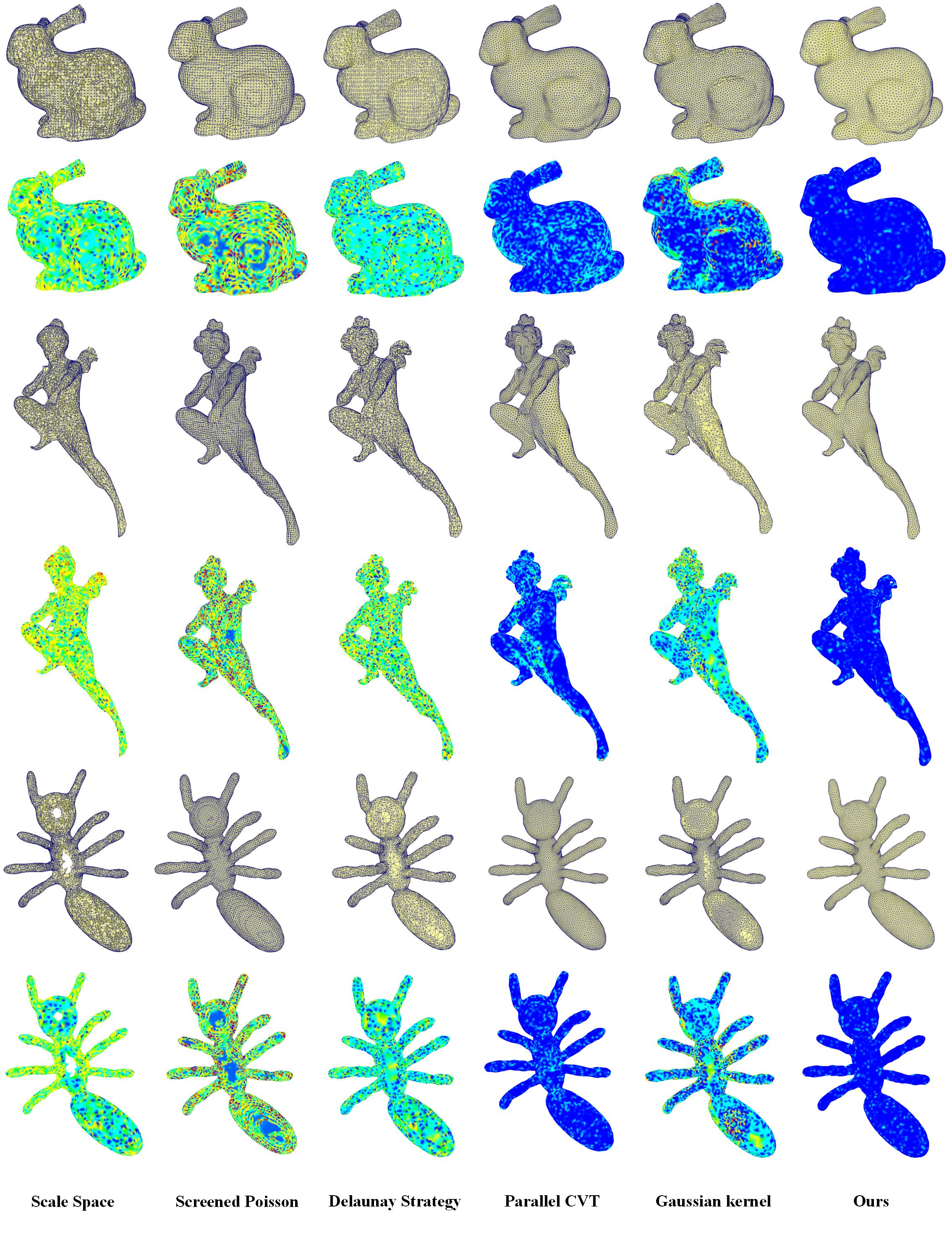}
  \caption{Comparisons of the reconstructed meshes and their color maps by different methods. The point clouds from top to bottom are Bunny, Angle, and Ant.}
  \label{f15}
\end{figure*}

\begin{figure*}
  \includegraphics[width=\linewidth]{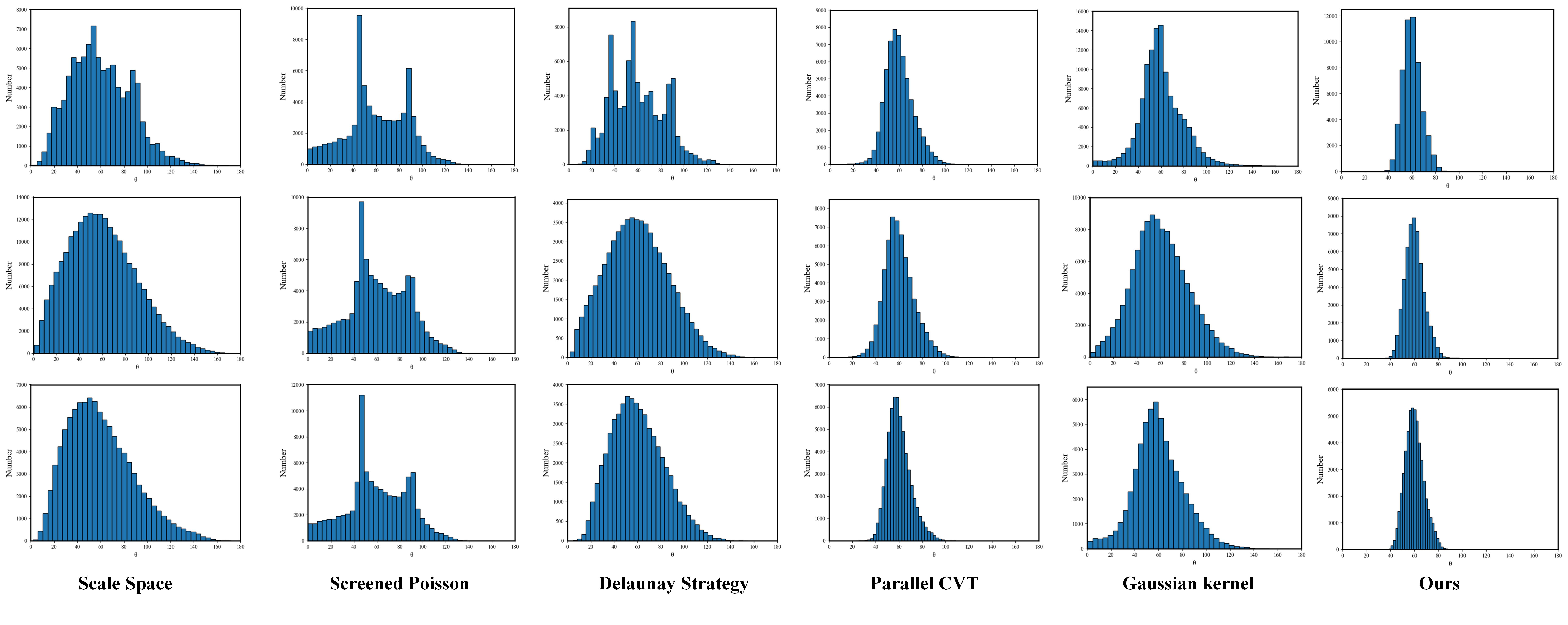}
  \caption{Comparisons of the $\theta$ histograms of different methods. The point clouds from top to bottom are Bunny, Angle, and Ant.}
  \label{f16}
\end{figure*}

\begin{figure*}[!htb]
  \center
  \includegraphics[scale=0.3]{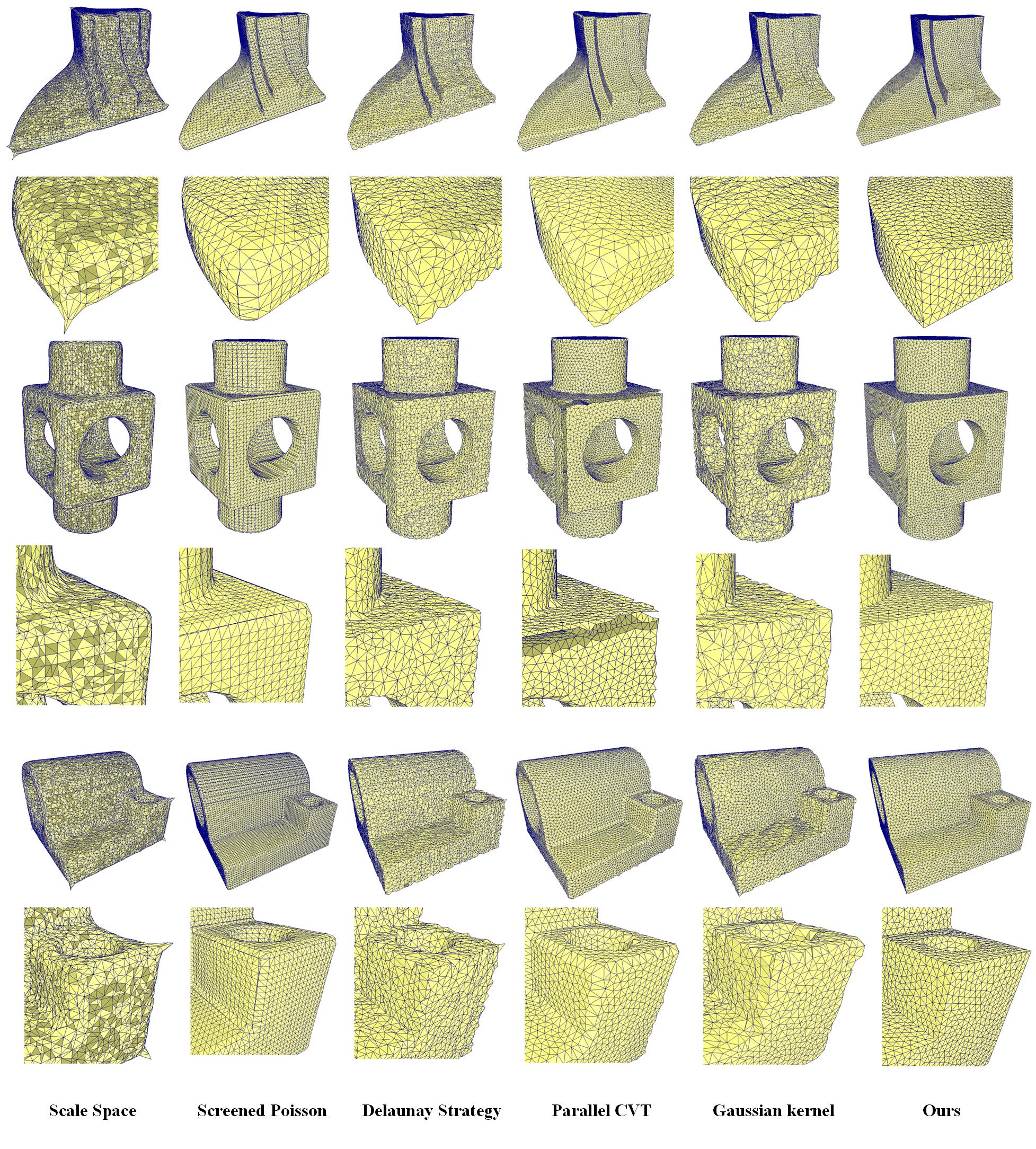}
  \caption{Comparisons of the reconstructed meshes with external edges by different methods. The point clouds from top to bottom are Fandisk, Block, and Joint.}
  \label{f17}
\end{figure*}

\begin{figure*}[!htb]
  \includegraphics[width=\linewidth]{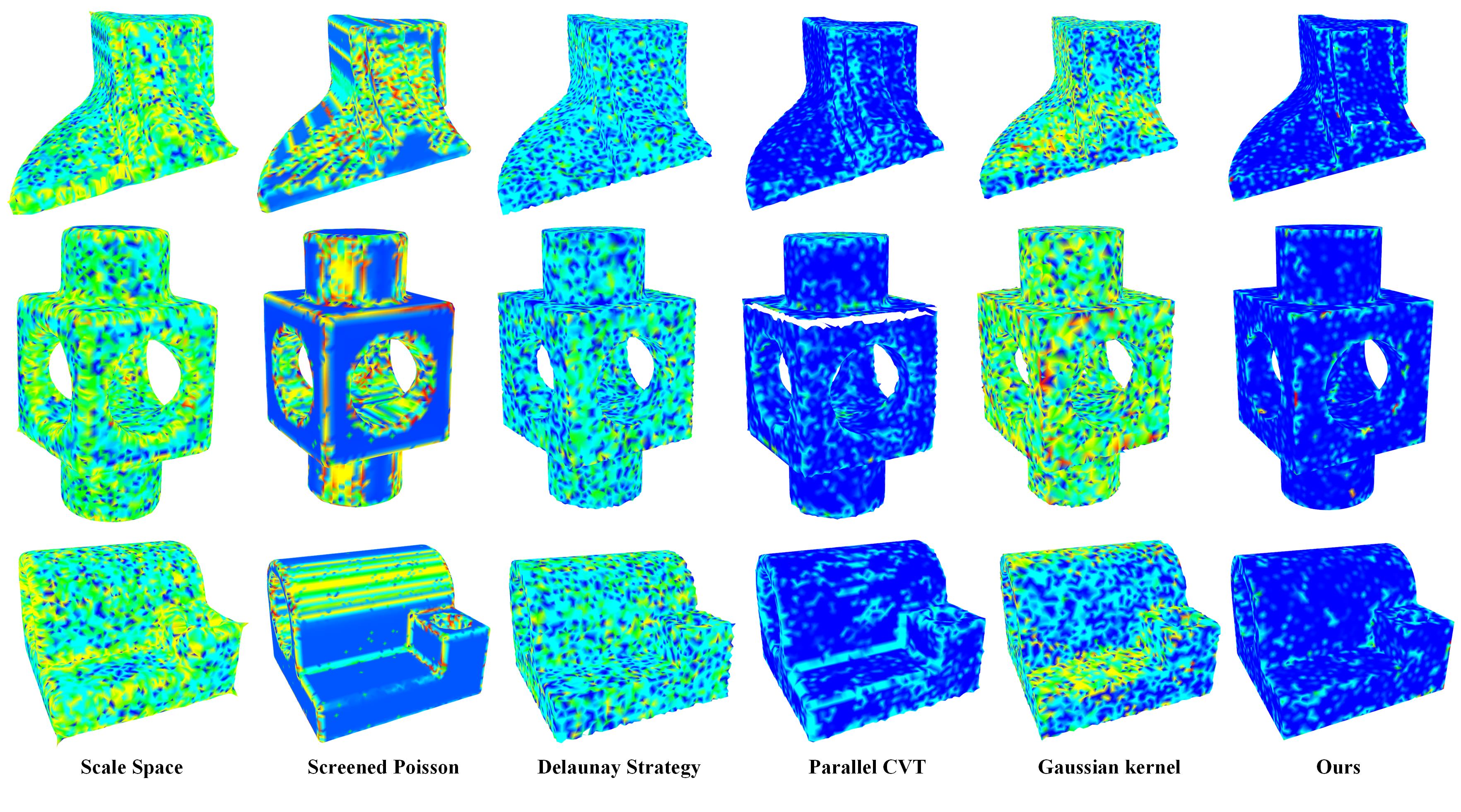}
  \caption{Comparisons of the color maps by different methods. The point clouds from top to bottom are Fandisk, Block, and Joint.}
  \label{f171}
\end{figure*}

\begin{figure*}[!htb]
  \includegraphics[width=\linewidth]{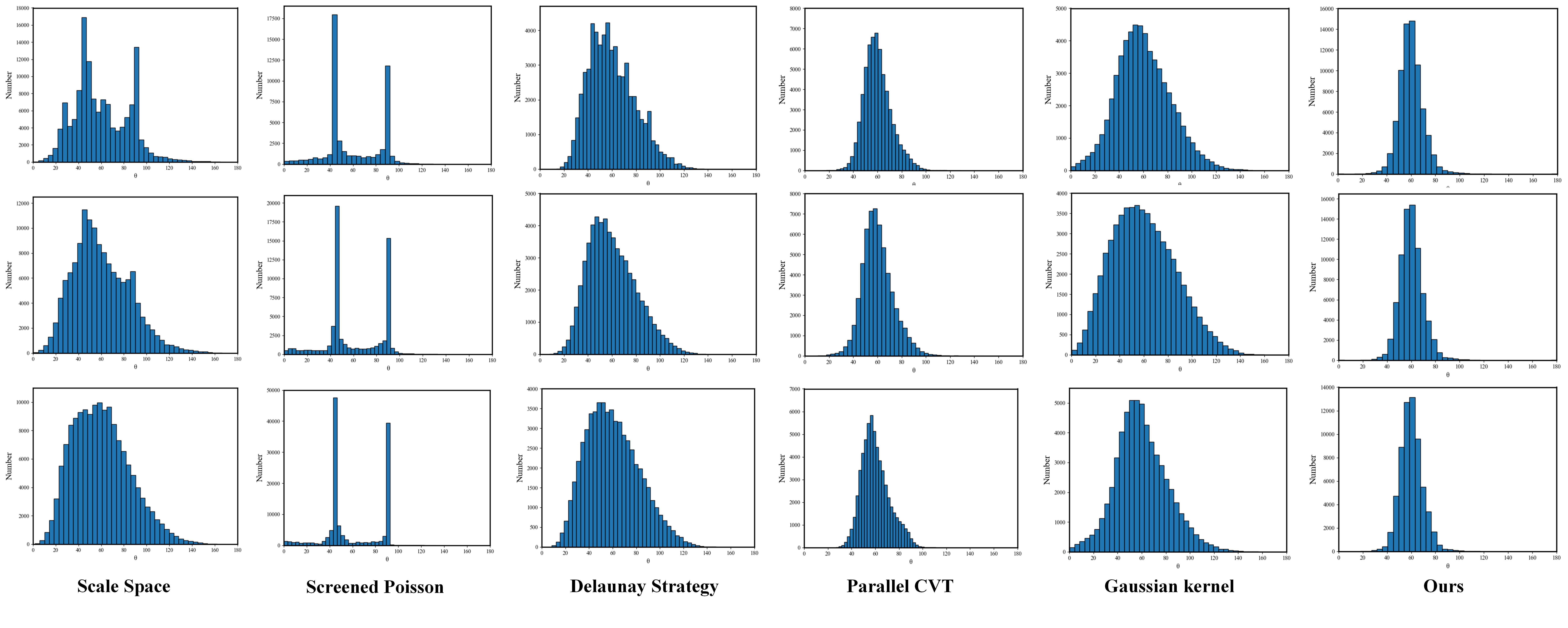}
  \caption{Comparisons of the $\theta$ histograms of different methods. The point clouds from top to bottom are Fandisk, Block, and Joint.}
  \label{f172}
\end{figure*}

\begin{table*}[]
\centering
\caption{Comparisons of minimum angle statistics by the different methods.}
\label{TN1}
\begin{tabular}{ccccccccccccc}
\hline
\textbf{Mehtod}                     & \multicolumn{2}{c}{\textbf{Scale Space}} & \multicolumn{2}{c}{\textbf{Screened Poisson}} & \multicolumn{2}{c}{\textbf{Delaunay Strategy}} & \multicolumn{2}{c}{\textbf{Parallel CVT}}  & \multicolumn{2}{c}{\textbf{Gaussian kernel}} & \multicolumn{2}{c}{\textbf{Ours}} \\ \hline
\textbf{Model(Point number)} & \textbf{$\theta_{min}$}       & \textbf{$\theta_{avg}$}       & \textbf{$\theta_{min}$}      & \textbf{$\theta_{avg}$}     & \textbf{$\theta_{min}$}          & \textbf{$\theta_{avg}$}         & \textbf{$\theta_{min}$} & \textbf{$\theta_{avg}$} & \textbf{$\theta_{min}$}      & \textbf{$\theta_{avg}$}      & \textbf{$\theta_{min}$} & \textbf{$\theta_{avg}$} \\ \hline
Bunny(10,000)                              & Nan                   & $34.73^\circ$               & Nan                  & $32.75^\circ$           & Nan                      & $36.84^\circ$                & $11.09^\circ$         & 48.87$^\circ$         & Nan                  & 43.61$^\circ$             & \textbf{36.31$^\circ$}         & \textbf{52.36$^\circ$}         \\
Horse(10,000)                               & Nan                   & 35.29$^\circ$               & Nan                  & 32.94$^\circ$             & Nan                      & 36.62$^\circ$                 & 8.182$^\circ$         & 49.05$^\circ$         & Nan                  & 38.80$^\circ$              & \textbf{35.85$^\circ$}         & \textbf{52.33$^\circ$}         \\
Armadillo(10,000)                           & Nan                   & 31.95$^\circ$               & Nan                  & 32.26$^\circ$              & Nan                      & 33.48$^\circ$                  & 9.68$^\circ$          & 49.29$^\circ$          & Nan                  & 39.66$^\circ$               & \textbf{27.03$^\circ$}          & \textbf{52.34$^\circ$}          \\
Dragon(20,000)                              & Nan                   & 32.06$^\circ$                & Nan                  & 32.50$^\circ$              & Nan                      & 34.27$^\circ$                 & Nan             & 48.84$^\circ$          & Nan                  & 38.81$^\circ$               & \textbf{17.96$^\circ$}          & \textbf{52.34$^\circ$}          \\
Asian Dragon(20,000)                        & Nan                   & 28.61$^\circ$                & Nan                  & 32.49$^\circ$              & Nan                      & 34.40$^\circ$                  & Nan             & 48.25$^\circ$          & Nan                  & 41.25$^\circ$               & \textbf{9.23$^\circ$}          & \textbf{52.20$^\circ$}          \\
Angel(10,000)                               & Nan                   & 31.11$^\circ$                & Nan                  & 32.03$^\circ$              & Nan                      & 32.94$^\circ$                  & 9.01$^\circ$          & 48.03$^\circ$          & Nan                  & 37.04$^\circ$               & \textbf{10.16$^\circ$}          & \textbf{52.31$^\circ$}          \\
Buddha(20,000)                              & Nan                   & 31.82$^\circ$                & Nan                  & 33.21$^\circ$              & Nan                      & 34.01$^\circ$                  & Nan             & 47.09$^\circ$          & Nan                  & 38.95$^\circ$               & \textbf{13.14$^\circ$}          & \textbf{52.29$^\circ$}          \\
Lucy(20,000)                                & Nan                   & 31.44$^\circ$                & Nan                  & 32.77$^\circ$              & Nan                      & 35.46$^\circ$                  & 6.77$^\circ$           & 47.84$^\circ$          & Nan                  & 36.29$^\circ$               & \textbf{10.38$^\circ$}          & \textbf{52.29$^\circ$}          \\
Thai Statue(20,000)                         & Nan                   & 29.81$^\circ$                & Nan                  & 32.63$^\circ$              & Nan                      & 34.33$^\circ$                  & 6.01$^\circ$          & 47.23$^\circ$          & Nan                  & 41.24$^\circ$                & \textbf{10.54$^\circ$}          & \textbf{52.07$^\circ$}          \\
Ant(10,000)                                 & Nan                   & 34.28$^\circ$               & Nan                  & 32.53$^\circ$              & Nan                      & 37.96$^\circ$                  & 23.11$^\circ$          & 50.91$^\circ$          & Nan                  & 40.37$^\circ$               & \textbf{30.71$^\circ$}          & \textbf{52.32$^\circ$}          \\
Centuar(10,000)                             & Nan                   & 34.42$^\circ$                & Nan                  & 32.32$^\circ$              & Nan                      & 38.67$^\circ$                  & 20.35$^\circ$          & 50.18$^\circ$          & Nan                  & 37.61$^\circ$               & \textbf{34.96$^\circ$}          & \textbf{52.35$^\circ$}          \\
Giraffe(10,000)                             & Nan                   & 33.52$^\circ$                & Nan                  & 33.27$^\circ$              & Nan                      & 37.09$^\circ$                  & 14.75$^\circ$          & 50.42$^\circ$         & Nan                  & 38.05$^\circ$               & \textbf{34.95$^\circ$}         & \textbf{52.33$^\circ$}          \\
Hand(10,000)                                & Nan                   & 36.07$^\circ$                & Nan                  & 32.59$^\circ$              & Nan                      & 38.44$^\circ$                  & 18.28$^\circ$          & 50.51$^\circ$          & Nan                  & 43.90$^\circ$               & \textbf{33.77$^\circ$}          & \textbf{52.33$^\circ$}          \\
Nunchakus(10,000)                           & Nan                   & 33.57$^\circ$                 & Nan                  & 33.10$^\circ$              & 7.25$^\circ$                   & 37.01$^\circ$                  & \textbf{14.09$^\circ$}          & 49.99$^\circ$          & Nan                  & 40.13$^\circ$               & 10.48$^\circ$           & \textbf{51.88$^\circ$}          \\
WoodMan(10,000)                             & Nan                   & 32.46$^\circ$                & Nan                  & 32.58$^\circ$              & Nan                      & 35.56$^\circ$                  & 10.51$^\circ$          & 50.43$^\circ$          & Nan                  & 38.44$^\circ$               & \textbf{32.30$^\circ$}          & \textbf{52.33$^\circ$}          \\
Block(10,000)                               & Nan                   & 36.84$^\circ$                & Nan                  & 36.21$^\circ$              &7.93$^\circ$                  & 41.33$^\circ$                  & 7.05$^\circ$          & 48.45$^\circ$          & Nan                  & 33.82$^\circ$               & \textbf{8.29$^\circ$}          & \textbf{51.09$^\circ$}          \\
Fandisk(10,000)                                & Nan                   & 37.18$^\circ$                & Nan                  & 37.70$^\circ$              &10.40$^\circ$                   & 42.02$^\circ$                  & \textbf{14.81$^\circ$}          & 49.79$^\circ$          & Nan                  & 38.56$^\circ$               & 8.12$^\circ$          & \textbf{51.06$^\circ$}           \\
Joint(10,000)                               & Nan                   & 36.12$^\circ$                 & Nan                  & 37.44$^\circ$              & 8.36$^\circ$                  & 38.38$^\circ$                  & \textbf{26.59$^\circ$}          & 49.56$^\circ$          & Nan                  & 39.82$^\circ$               & 13.96$^\circ$           & \textbf{51.28$^\circ$}    \\ \hline
\end{tabular}
\end{table*}

\begin{table*}[]
\centering
\caption{Comparisons of triangle quality measurement by the different methods.}
\label{TN2}
\begin{tabular}{ccccccccccccc}
\hline
\textbf{Method} & \multicolumn{2}{c}{\textbf{Scale Space}} & \multicolumn{2}{c}{\textbf{Screened Poisson}} & \multicolumn{2}{c}{\textbf{Delaunay Strategy}} & \multicolumn{2}{c}{\textbf{Parallel  CVT}}  & \multicolumn{2}{c}{\textbf{Gaussian kernel}} & \multicolumn{2}{c}{\textbf{Ours}} \\\hline
\textbf{Model(Point number)}  & \textbf{$Q_{min}$}       & \textbf{$Q_{avg}$}       & \textbf{$Q_{min}$}      & \textbf{$Q_{avg}$}     & \textbf{$Q_{min}$}          & \textbf{$Q_{avg}$}         & \textbf{$Q_{min}$} & \textbf{$Q_{avg}$} & \textbf{$Q_{min}$}      & \textbf{$Q_{avg}$}      & \textbf{$Q_{min}$} & \textbf{$Q_{avg}$} \\\hline
Bunny(10,000)           & Nan                   & 0.66                  & Nan                  & 0.61                & 0.11                     & 0.70                     & 0.19            & 0.86            & Nan                  & 0.77                 & \textbf{0.59}   & \textbf{0.91}   \\
Horse(10,000)           & Nan                   & 0.67                  & Nan                  & 0.61                & 0.12                     & 0.69                    & 0.20             & 0.86            & Nan                  & 0.72                 & \textbf{0.62}   & \textbf{0.91}   \\
Armadillo(10,000)       & Nan                   & 0.62                  & Nan                  & 0.61                & 0.14                     & 0.65                    & 0.26            & 0.86            & Nan                  & 0.73                 & \textbf{0.47}   & \textbf{0.91}   \\
Dragon(20,000)          & Nan                   & 0.62                  & Nan                  & 0.61                & Nan                      & 0.66                    & Nan             & 0.86            & Nan                  & 0.72                 & \textbf{0.40}    & \textbf{0.91}   \\
Asian Dragon(20,000)    & Nan                   & 0.56                  & Nan                  & 0.61                & 0.11                     & 0.66                    & Nan             & 0.85            & Nan                  & 0.75                 & \textbf{0.19}   & \textbf{0.91}   \\
Angel(10,000)           & Nan                   & 0.60                   & Nan                  & 0.60                 & Nan                      & 0.64                    & 0.19            & 0.85            & Nan                  & 0.69                 & \textbf{0.26}   & \textbf{0.91}   \\
Buddha(20,000)          & Nan                   & 0.61                  & Nan                  & 0.62                & Nan                      & 0.65                    & Nan             & 0.84            & Nan                  & 0.72                 & \textbf{0.28}   & \textbf{0.91}   \\
Lucy(20,000)            & Nan                   & 0.61                  & Nan                  & 0.62                & Nan                      & 0.68                    & 0.16            & 0.85            & Nan                  & 0.69                 & \textbf{0.22}   & \textbf{0.91}   \\
Thai Statue(20,000)     & Nan                   & 0.58                  & Nan                  & 0.61                & Nan                      & 0.66                    & 0.16            & 0.84            & Nan                  & 0.75                 & \textbf{0.22}   & \textbf{0.90}    \\
Ant(10,000)             & Nan                   & 0.65                  & Nan                  & 0.61                & 0.13                     & 0.71                    & 0.45            & 0.88            & Nan                  & 0.74                 & \textbf{0.56}   & \textbf{0.91}   \\
Centuar(10,000)         & Nan                   & 0.65                  & Nan                  & 0.61                & 0.11                     & 0.72                    & 0.42            & 0.87            & Nan                  & 0.71                 & \textbf{0.59}   & \textbf{0.91}   \\
Giraffe(10,000)         & Nan                   & 0.64                  & Nan                  & 0.62                & 0.13                     & 0.70                     & 0.33            & 0.88            & Nan                  & 0.71                 & \textbf{0.64}   & \textbf{0.91}   \\
Hand(10,000)            & Nan                   & 0.68                  & Nan                  & 0.61                & Nan                      & 0.72                    & 0.38            & 0.88            & Nan                  & 0.78                 & \textbf{0.57}   & \textbf{0.91}   \\
Nunchakus(10,000)       & Nan                   & 0.64                  & Nan                  & 0.62                & 0.18                     & 0.70                     & \textbf{0.32}   & 0.87            & Nan                  & 0.74                 & 0.18            & \textbf{0.90}    \\
WoodMan(10,000)         & Nan                   & 0.62                  & Nan                  & 0.61                & Nan                      & 0.68                    & 0.22            & 0.88            & Nan                  & 0.72                 & \textbf{0.54}            & \textbf{0.91}   \\
Block(10,000)           & Nan                   & 0.69                  & Nan                  & 0.63                & 0.22                     & 0.75                    & 0.18            & 0.85            & Nan                  & 0.65                 & \textbf{0.18}   & \textbf{0.89}   \\
Fandisk(10,000)            & Nan                   & 0.68                  & Nan                  & 0.65                & 0.27                     & 0.76                    & \textbf{0.38}   & 0.87            & Nan                  & 0.72                 & 0.17            & \textbf{0.89}   \\
Joint(10,000)           & Nan                   & 0.68                  & Nan                  & 0.64                & 0.23                     & 0.72                    & \textbf{0.44}   & 0.86            & Nan                  & 0.73                 & 0.28            & \textbf{0.90} \\ \hline
\end{tabular}
\end{table*}

\subsection{Evaluations}

To measure the quality of the reconstructed mesh, we introduce some evaluation metrics including triangle quality measurement \cite{frey1999surface}, minimum angle statistics, and visualization. The triangle quality measurement $Q(t)\in\lbrack0,1\rbrack$ evaluates the isotropic property of a triangle $t$ from the reconstructed mesh. The minimum $Q_{min}$ and average $Q_{ave}$ are computed from $Q(t)$ to quantify the isotropic property. Another important evaluation metric is the minimum interior angle degree $\theta(t)$ of a triangle. The high-quality triangular mesh should avoid too large or small angles \cite{wang2018isotropic}. Similar to $Q_{min}$ and $Q_{ave}$, $\theta_{min}$ and $\theta_{ave}$ are used for evaluation. Using histogram and color map to visualize the statistics of $\{\theta(t)\}$, the quality of the reconstructed mesh is shown clearly.

For the histogram visualization, we count all angle values to generate the histogram. For color map visualization, we compute the minimum angle of each vertex and obtain the color value by color transfer. The minimum angle of a vertex is computed by the following steps: 1. collecting the triangles $\{t_{p}\}$ that include the vertex $p$; 2. computing the angels $\{\theta_{p}\}$ which take $p$ as an angle vertex; 3. transferring the angel value to color value. In Figure \ref{f14}, we show an instance of the color map and histogram from a reconstructed mesh.

\subsection{Comparisons}

Based on the evaluation metrics, we compare the performance of different reconstruct methods including Scale Space \cite{Digne2011Scale}, Screened Poisson \cite{kazhdan2013screened}, Delaunay Strategy \cite{cohen2004greedy} \cite{Lafarge2013surface}, Parallel CVT \cite{Chen2018CVT}, and Gaussian kernel-based optimization \cite{zhong2019surface}. The Scale Space and Delaunay Strategy are implemented with CGAL library (www.cgal.org). The Screened Poisson result is achieved from MeshLab toolbox(www.meshlab.net). The Parallel CVT and Gaussian kernel-based optimization are programmed in our platform with the assistant of CGAL library. The iteration numbers for Parallel CVT is set to 50. For L-BFGS implementation of Gaussian kernel-based optimization, a hybrid L-BFGS tool (HLBFGS) is used (xueyuhanlang.github.io/software/HLBFGS). In our framework, we set the iteration step to 5 for isotropic remeshing.

The test set contains different kinds of point clouds such as dense point clouds, sparse point clouds, and point clouds with external edges. It can be used to evaluate the performance of different reconstruction methods comprehensively. The resampling point number is set to 10,000 or 20,000. As some methods can not control the resampling point number, a point cloud resampling should be added as an assistant. The
resampling method is provided by CGAL library (grid-based method). In Figure \ref{f15}, we show the reconstructed meshes and their color maps obtained with different methods. Our method achieves better isotropic property in reconstructed meshes. The related $\theta$ histogram results are shown in Figure \ref{f16}. In Table \ref{TN1} and \ref{TN2}, we show the comparisons of $\theta$ and $Q$ for the test set. If $Q$ is less than 0.1 or the $\theta$ is less than $5^\circ$, it means the related triangle $t$ is a low-quality one. We label the value to "Nan"(non-acceptable value). The comparative data supports the conclusion that our method achieves better meshes. To further demonstrate the performance of external edge keeping, we compare the reconstructed meshes with external edges by different methods in Figures \ref{f17}-\ref{f172}. It shows that our method remains the accurate external edges while keeping the isotropic property. It keeps the 2-manifold property and avoids the wrong truncation (such as the Block model by Parallel CVT).

\begin{figure}
  \includegraphics[width=\linewidth]{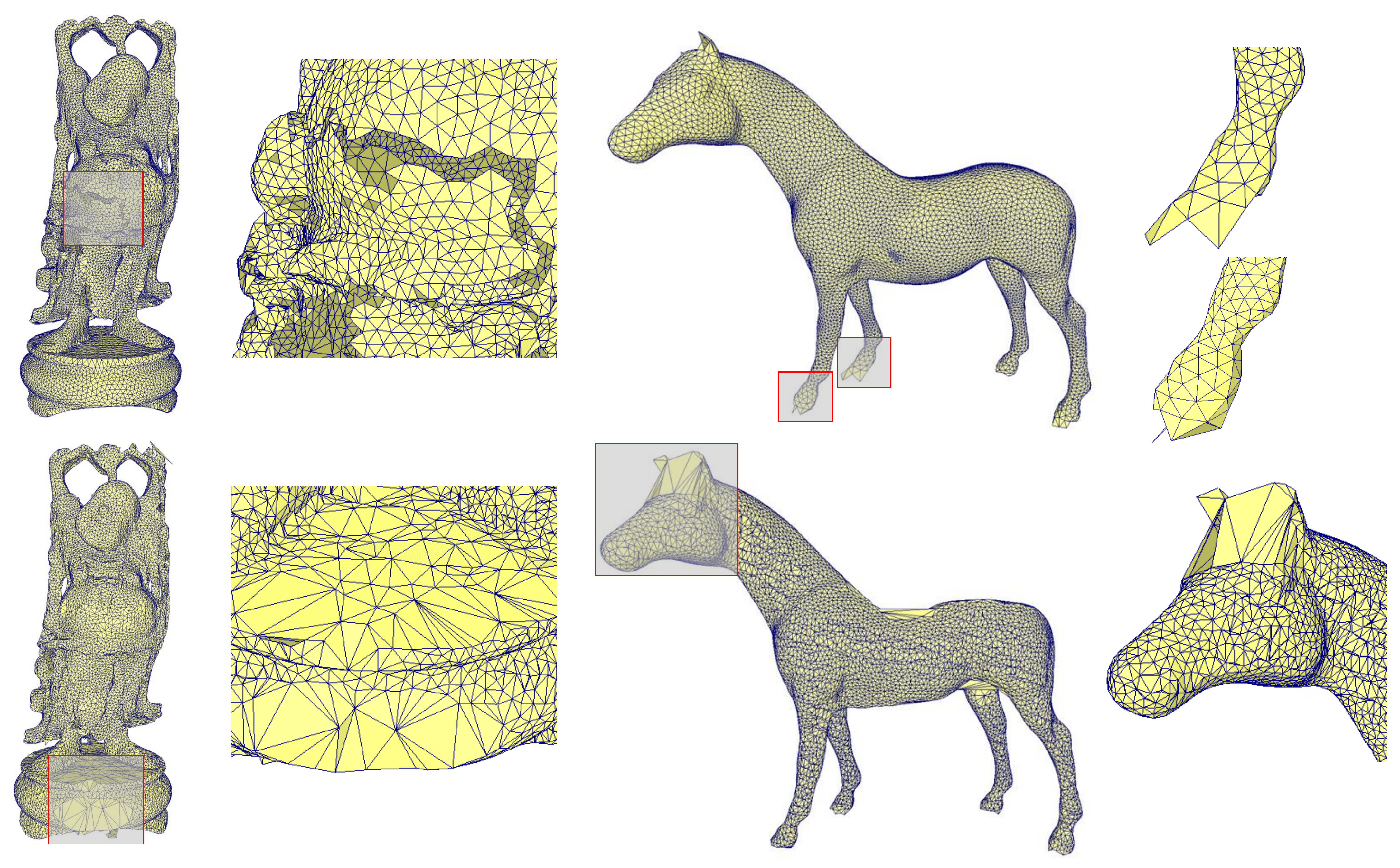}
  \caption{Instances of the incorrect triangles in reconstruct meshes (Buddha and Horse). The top row: reconstruct meshes by Parallel CVT; the bottom row: reconstruct meshes by Gaussian kernel-based optimization.}
  \label{f18}
\end{figure}

\begin{figure}
  \includegraphics[width=\linewidth]{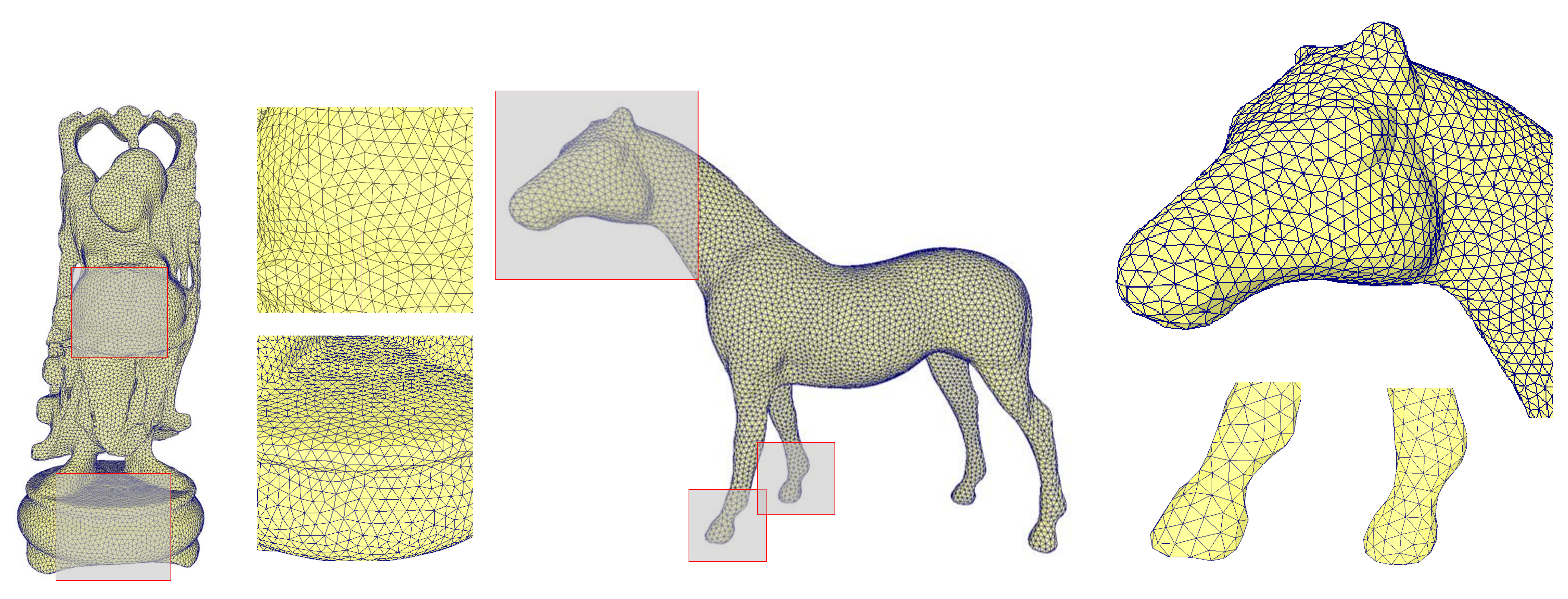}
  \caption{Instances of our reconstructed meshes (Buddha and Horse). Based on the intrinsic metric, the incorrect triangles are removed.}
  \label{f19}
\end{figure}

\begin{figure}
  \includegraphics[width=\linewidth]{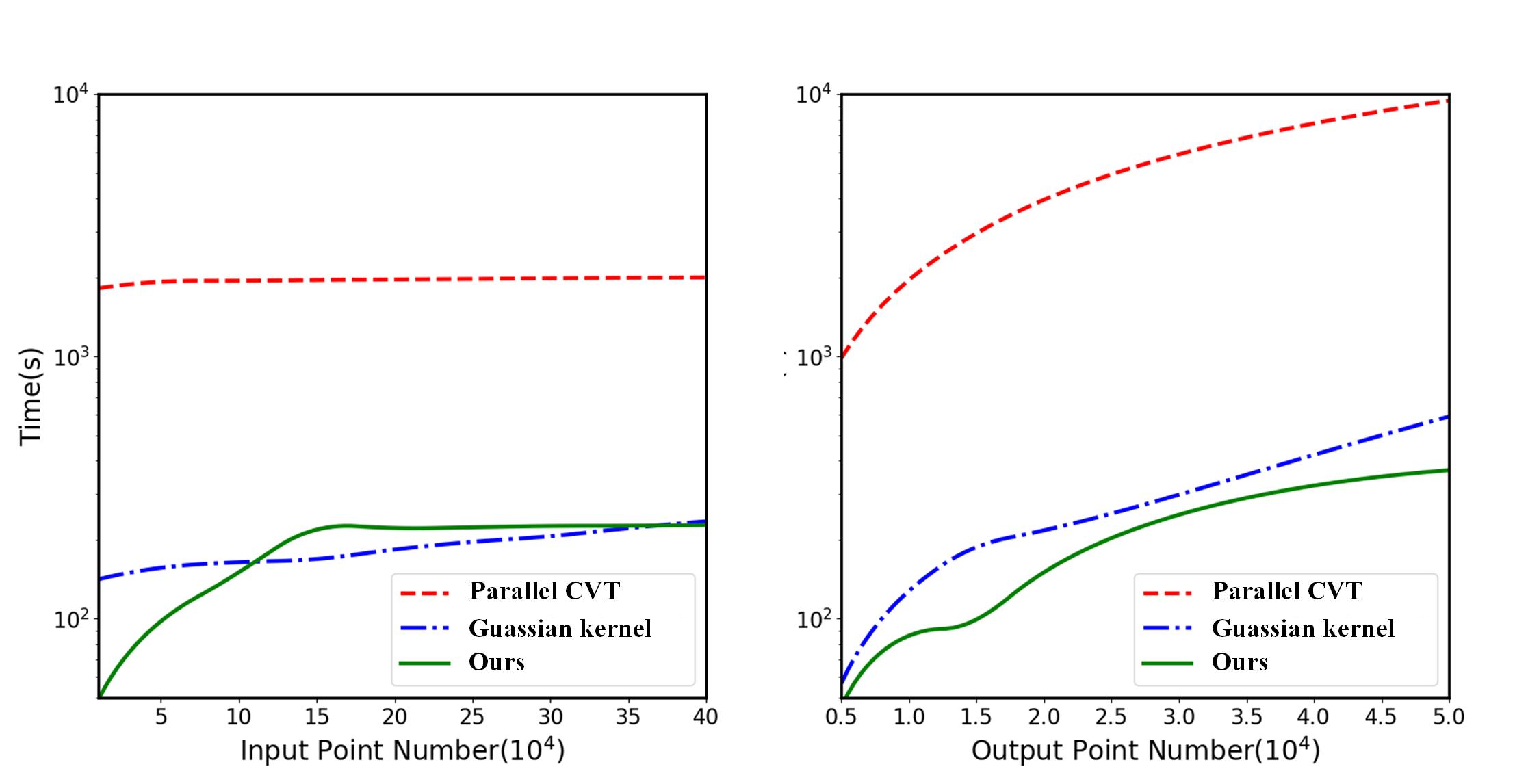}
  \caption{The time cost of three methods.}
  \label{f20}
\end{figure}

\begin{figure}
  \includegraphics[width=\linewidth]{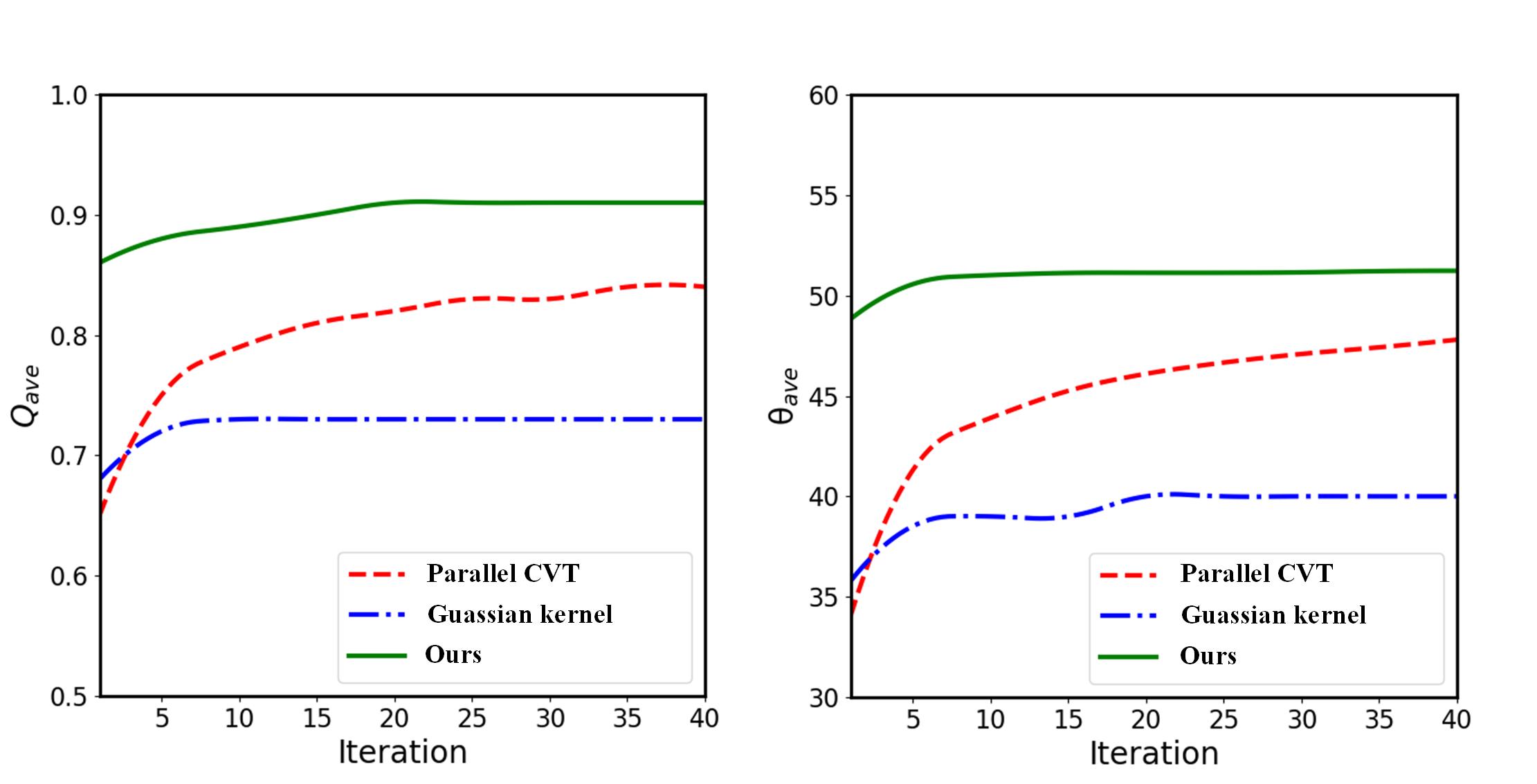}
  \caption{The convergence reports ($Q(t)$ and $\theta(t)$) of the three methods.}
  \label{f21}
\end{figure}

\subsection{Analysis}

In this part, we make a further analysis for different reconstruct methods. The Scale Space is limited by the density of the point cloud. Once the point cloud has some regions with non-uniform point density, the reconstructed mesh is not satisfied (such as the Ant model in Figure \ref{f15}). Using the Screened Poisson to reconstruct mesh, the 2-manifold property can be kept. The reason is the reconstruction is not limited by the local tangent space. The reconstruction is processed in a global optimization scheme. The advantage of the method is running faster with 2-manifold property.
The drawbacks of the method are the quality of the mesh is not favorable and some geometric features are lost. The Delaunay Strategy harvests the reconstructed mesh without point movement. The core of many reconstruction methods is based on the strategy. However, it is limited by the density of point distribution. The above three methods cannot precisely control the point number in the reconstructed mesh.

As the state of the art, the Parallel CVT and Gaussian kernel-based optimization are capable of controlling the point number in the reconstructed mesh while keeping the isotropic property. However, as aforementioned, such methods lost some geometric features after points' smoothing in the local tangent space. If the user-specified resampling point number is too small, the point density is reduced. Then the distance between two adjacent points in tangent space can not be used to represent the real point distance on the mesh, which affects the accuracy of local region detection, especially for the regions with sharp curvature changes. Some instances are shown in Figure \ref{f18}. In our framework, such errors are avoided by the voxel structure. In Figure \ref{f19}, we show the improvement of our framework.

For time cost analysis, we compare two kinds of mesh reconstruction tasks: 1. input point clouds with different point number and output the certain resampling point number (10,000) in the reconstructed meshes; 2. fix the input point number (200,000) and output different resampling point numbers in the reconstructed meshes. The time overhead of the three methods is shown in Figure \ref{f20}. For Parallel CVT, it is not sensitive to the input point number. However, it is sensitive to the output resampling point number. The time cost of the Voronoi diagram is larger than other methods even the computation can be accelerated by a multi-core structure. Thanks to the acceleration of HLBFGS, the Gaussian kernel-based optimization speed is faster than Parallel CVT. It is not sensitive to the output or input point number. Our method achieves similar performance to the Gaussian kernel-based optimization.

\begin{table}[]
\centering
\scriptsize
\caption{Comparisons of MLS error reports by the different methods.}
\label{TN3}
\begin{tabular}{ccccccc}
\hline
\textbf{Methods}                                                 & \multicolumn{2}{c}{\textbf{Parallel CVT}}      & \multicolumn{2}{c}{\textbf{Particle-based}} & \multicolumn{2}{c}{\textbf{Ours}}     \\ \hline
\textbf{Model}                                                   & \textbf{$\Delta$Max}      & \textbf{$\Delta$Avg}      & \textbf{$\Delta$Max}         & \textbf{$\Delta$Avg}         & \textbf{$\Delta$Max}      & \textbf{$\Delta$Avg}      \\ \hline
\textbf{Bunny}                                                   & 2.65E-2           & 4.12E-3          & 4.12E-1               & 1.46E-2              & \textbf{2.02E-2}  & \textbf{2.96E-3} \\
\textbf{Horse}                                                   & Nan               & Nan               & 1.08E-1               & 5.24E-3             & \textbf{1.87E-2}  & \textbf{1.75E-3} \\
\textbf{Armadillo}                                               & 2.54E-2           & \textbf{3.42E-3} & 2.51E-2              & 3.54E-3             & \textbf{2.27E-2}  & 4.68E-3          \\
\textbf{Dragon}                                                  & 3.24E-2            & \textbf{2.74E-3} & 9.33E-2              & 3.27E-2              & \textbf{2.82E-2}  & 5.04E-3          \\
\textbf{\begin{tabular}[c]{@{}c@{}}Asian \\ Dragon\end{tabular}} & \textbf{1.94E-2}  & \textbf{2.33E-3} & 2.10E-2              & 3.62E-3             & 2.02E-2           & 4.13E-3          \\
\textbf{Angel}                                                   & 1.95E-2           & 2.94E-3          & 4.06E-2              & \textbf{2.76E-3}    & \textbf{1.94E-2}  & 2.99E-3          \\
\textbf{Buddha}                                                  & 2.04E-2           & 3.20E-3          &5.18E-2              & 6.24E-3              & \textbf{1.81E-2}  & \textbf{3.10E-3} \\
\textbf{Lucy}                                                    & 2.39E-2           & 3.54E-3          & \textbf{1.49E-2}     & 2.62E-3             & 1.79E-2           & \textbf{3.18E-3} \\
\textbf{\begin{tabular}[c]{@{}c@{}}Thai \\ Statue\end{tabular}}  & Nan               & Nan               & 6.33E-2              & 9.28E-3             & \textbf{2.01E-2}  & \textbf{3.77E-3} \\
\textbf{Ant}                                                     & 8.83E-3          & 2.35E-3          & 1.50E-2              & 2.28E-3              & \textbf{8.01E-3} & \textbf{1.61E-3} \\
\textbf{Centuar}                                                 & \textbf{1.51E-2}  & \textbf{2.96E-3} & 2.08E-2              & 4.25E-3             & 2.06E-2           & 2.98E-3          \\
\textbf{Giraffe}                                                 & \textbf{1.52E-2}  & 2.12E-3          & 2.25E-2              & 4.07E-3             & 2.19E-2           & \textbf{1.56E-3} \\
\textbf{Hand}                                                    & \textbf{1.18E-2}  & 2.36E-3          & 1.25E-2              & 2.22E-3             & 1.67E-2           & \textbf{2.14E-3} \\
\textbf{Nunchakus}                                               & 1.44E-2           & 2.19E-3           & 3.35E-2              & 9.27E-3             & \textbf{1.05E-2}  & \textbf{1.58E-3} \\
\textbf{WoodMan}                                                 & \textbf{9.98E-3} & 2.53E-3          & 1.46E-2              & 3.29E-3             & 1.13E-2           & \textbf{2.23E-3} \\
\textbf{Block}                                                   & 4.47E-2           & 6.96E-3          & \textbf{3.39E-2}     & 3.96E-3             & 3.41E-2           & \textbf{3.95E-3} \\
\textbf{Fandisk}                                                    & 3.06E-2           & 2.82E-3          & 2.90E-2              & 3.03E-3             & \textbf{3.06E-2}  & \textbf{2.35E-3} \\
\textbf{Joint}                                                   & 5.48E-2           & 4.12E-3          & 5.86E-2              & 8.39E-3              & \textbf{4.99E-2}  & \textbf{3.07E-3} \\ \hline
\end{tabular}
\end{table}

In convergence analysis, we compare the $Q_{ave}$ and $\theta_{ave}$ in different iterations of three methods. The details are shown in Figure \ref{f21}. The Gaussian kernel-based optimization achieves faster convergence speed than Parallel CVT. However, it is sensitive to the initial resampling points‘ positions and is easy to fall into the local optimum. The values of $Q_{ave}$ and $\theta_{ave}$ are worse. For Parallel CVT, the convergence speed is slower. The values of $Q_{ave}$ and $\theta_{ave}$ are not converged even after 50 iterations. In contrast, our method achieves faster convergence speed with better values of $Q_{ave}$ and $\theta_{ave}$.

\begin{figure*}
  \includegraphics[width=\linewidth]{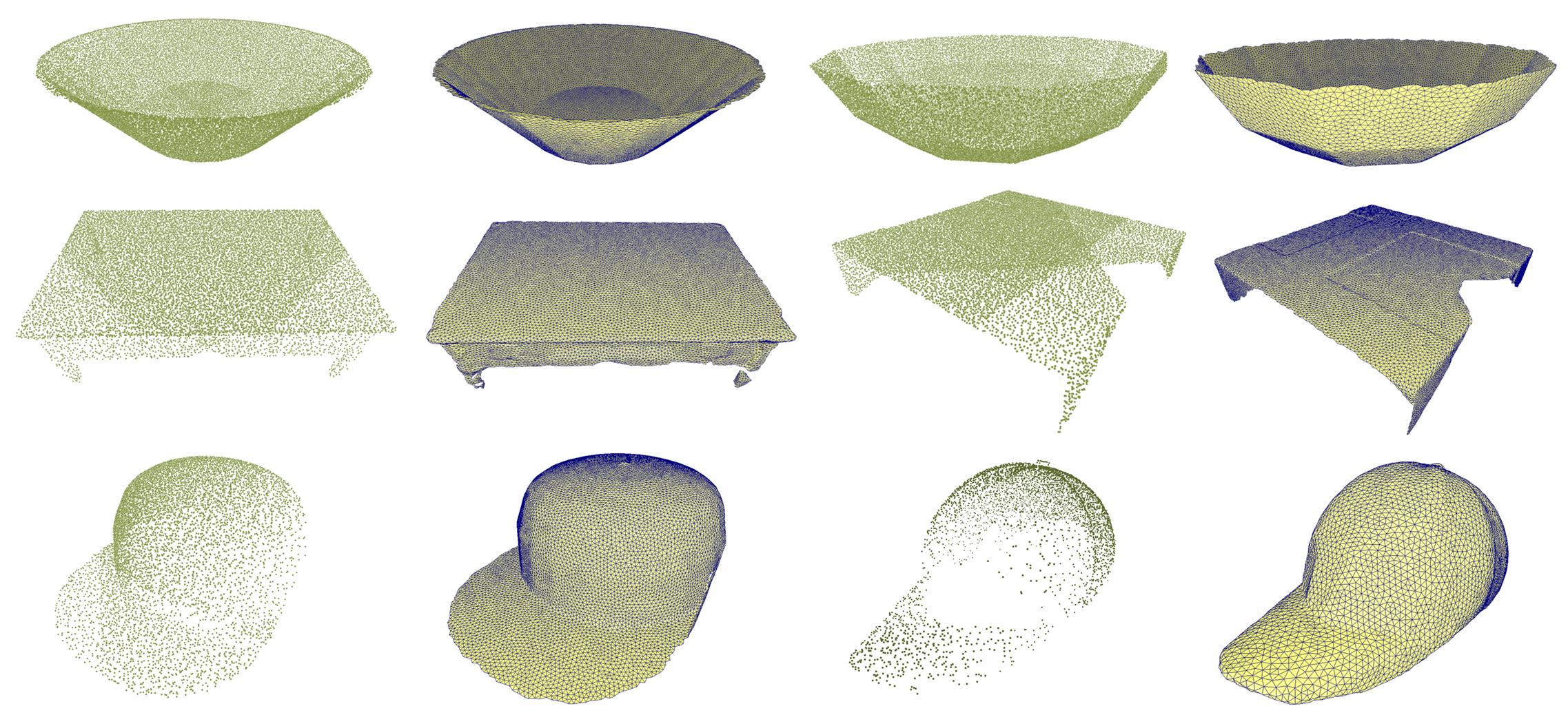}
  \caption{Instances of our mesh reconstruction in RGB-D Scenes Dataset.}
  \label{af1}
\end{figure*}

\begin{figure*}
  \includegraphics[width=\linewidth]{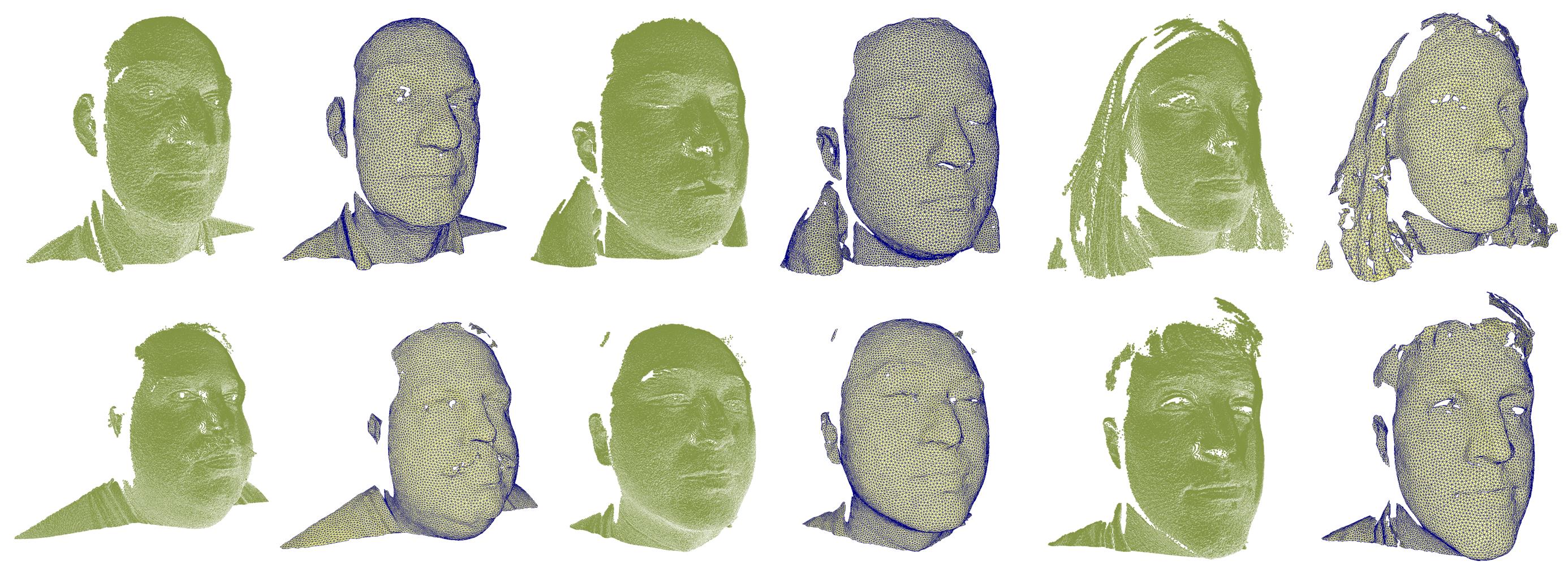}
  \caption{Instances of our mesh reconstruction in FRGC2.0.}
  \label{af2}
\end{figure*}

For geometric consistency evaluation, we compute the MLS error \cite{alexa2004normals} based on the state of the art and our method. To achieve the isotropic reconstructed mesh, some point positions are changed inevitably. If a reconstructed mesh maintains better geometric consistency to a point cloud, then the MLS error should be approach to zero. It means that the reconstructed mesh should fit the MLS surface well. We provide MLS error reports of Parallel CVT, Gaussian kernel-based optimization, and our framework in Table \ref{TN3}. MLS error results show that our mesh reconstruction achieves better performance for geometric consistency keeping.

\begin{figure*}
  \includegraphics[width=\linewidth]{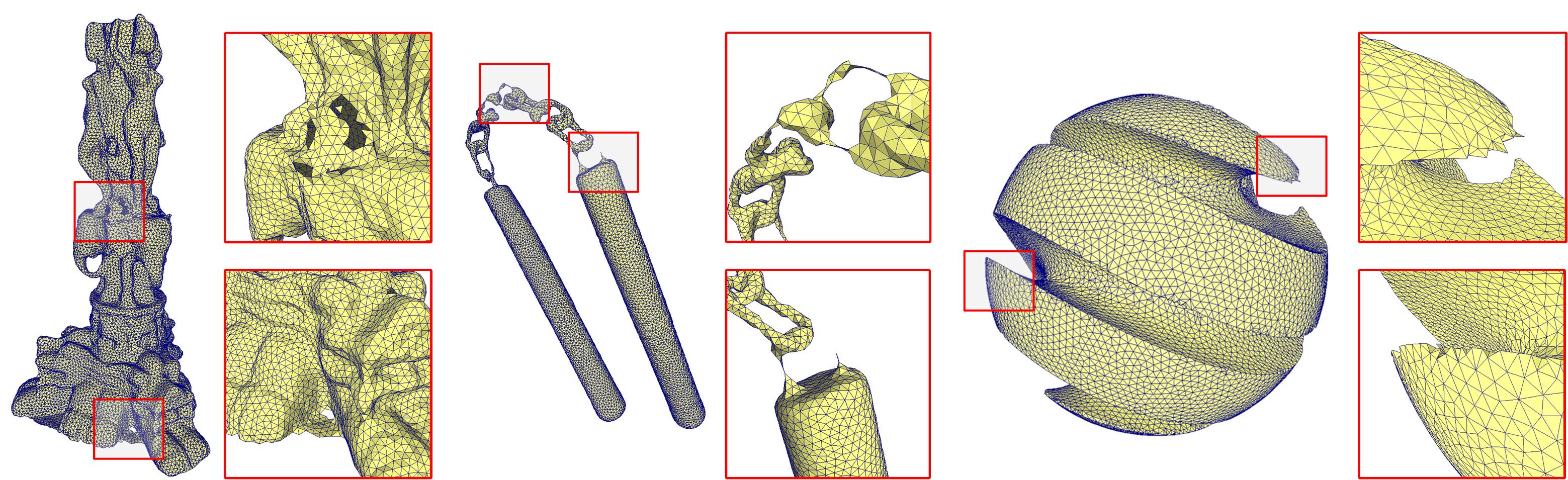}
  \caption{Some inaccurate instances of our mesh reconstruction in Stanford and SHREC.}
  \label{af3}
\end{figure*}

\begin{figure*}
  \includegraphics[width=\linewidth]{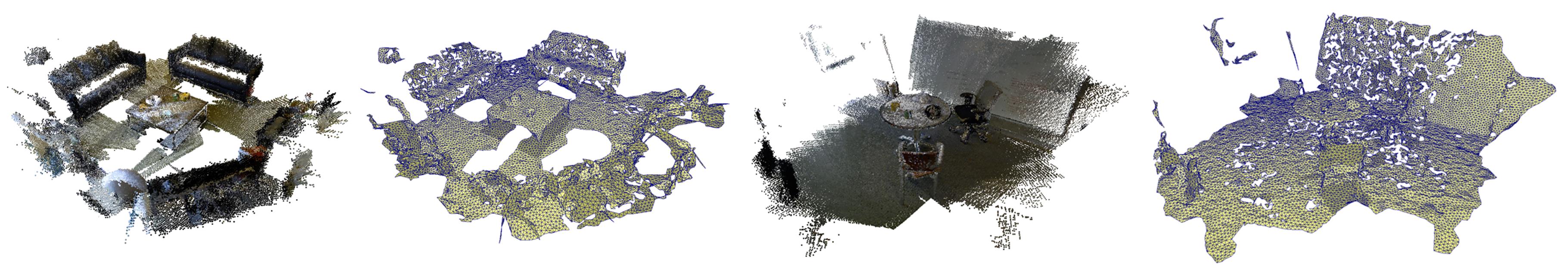}
  \caption{Some inaccurate instances of our mesh reconstruction in RGB-D Scenes Dataset.}
  \label{af4}
\end{figure*}

More mesh reconstruction results based on our framework are shown in Figures \ref{af1}-\ref{af4}. The additional test point clouds are collected from RGB-D Scenes Dataset and FRCG2.0. In Figures \ref{af1} and \ref{af2}, some reconstructed meshes from RGB-D Scenes Dataset and FRGC2.0 are shown. It proves that our framework has good performance for different datasets. However, to some point clouds, our framework can not achieve accurate reconstructed meshes. Such point clouds can be divided into two categories. One kind of point cloud has complex geometric features such as delicate texture, complicated topology structure, or extremely sharp edges (both external and internal, included angle is less than 30$^\circ$). Using our framework, such geometric details may be broken to a certain degree. In Figure \ref{af3}, some inaccurate reconstructed meshes are shown. Another kind of point cloud is not single object model. It contains a larger scene with multiple objects. For this kind of point cloud, our framework can not reconstruct the scene with accurate mesh. In Figure \ref{af4}, we show two instances. The reason is that our framework is designed for single object. For scene reconstruction, our framework does not consider the relationships between different objects.

\section{Conclusions}

We have introduced a voxel structure-based framework for mesh reconstruction. The proposed voxel structure provides the intrinsic metric which improves the accuracy of the initial reconstructed mesh. The point-based distance field is optimized by the voxel structure. It provides the correct point connections and maintains the user-specified point number. With different resampling rates for different point subsets, geometric features such as external edges and curvature sensitive property can be kept in the initial reconstructed mesh. Based on the initial reconstructed mesh, a mesh optimization method is proposed to further improve the mesh quality. The experimental results show that our framework achieves better reconstructed meshes. It obtains better isotropic property and keeps important geometric features. It is also shown that our framework achieves better convergence and faster speed.

\ifCLASSOPTIONcaptionsoff
  \newpage
\fi
\bibliographystyle{IEEEtran}
\bibliography{IEEEabrv,IEEEBib}

\begin{IEEEbiography}[{\includegraphics[width=1in,height=1.25in,clip,keepaspectratio]{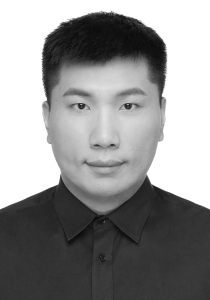}}]{Chenlei Lv}
received PhD degree in College of information science and technology, Beijing Normal University (BNU). He is currently a Post-doctor in School of Computer Science and Engineering, Nanyang Technology University (NTU). His research interests include Computer Vision, 3D Biometrics, Computer Graphics, Discrete Differential Geometry and Conformal Geometric. He has published several papers in Pattern Recognition, ACM Transactions on Multimedia Computing Communications and Applications, Pattern Recogniton Letter, etc. The personal page of the link is: \href{https://aliexken.github.io/}{https://aliexken.github.io/}.
\end{IEEEbiography}

\begin{IEEEbiography}[{\includegraphics[width=1in,height=1.25in,clip,keepaspectratio]{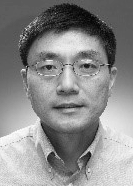}}]{Weisi Lin}
(M'92-SM'98-F'16) received the Ph.D. degree from King's College London, U.K. He is currently a Professor with the School of Computer Science and Engineering, Nanyang Technological University.
His research interests include image processing, perceptual signal modeling, video compression, and multimedia communication, in which he has
published over 200 journal papers, over 230 conference
papers, filed seven patents, and authored two
books. He is a fellow of the IET and an Honorary
Fellow of the Singapore Institute of Engineering
Technologists. He was the Technical Program Chair of the IEEE ICME
2013, PCM 2012, QoMEX 2014, and the IEEE VCIP 2017. He has been an
Invited/Panelist/Keynote/Tutorial Speaker at over 20 international conferences
and was a Distinguished Lecturer of the IEEE Circuits and Systems Society
from 2016 to 2017 and the Asia-Pacific Signal and Information Processing
Association (APSIPA) from 2012 to 2013. He has been an Associate Editor
of the IEEE TRANSACTIONS ON IMAGE PROCESSING, the IEEE TRANSACTIONS
ON CIRCUITS AND SYSTEMS FOR VIDEO TECHNOLOGY, the IEEE
TRANSACTIONS ON MULTIMEDIA, and the IEEE SIGNAL PROCESSING
LETTERS.
\end{IEEEbiography}

\begin{IEEEbiography}[{\includegraphics[width=1in,height=1.25in,clip,keepaspectratio]{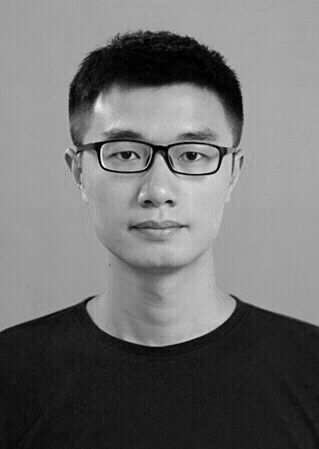}}]{Baoquan Zhao}
received his Ph.D. degree in computer science from Sun Yat-sen University, Guangzhou, China, in 2017. He is currently a Research Fellow with the School of Computer Science and Engineering, Nanyang Technological University, Singapore. He has served as a reviewer and technical program committee member of several journals and international conferences. He is a recipient of the Outstanding Reviewer Award of 2020 IEEE ICME. His research interests include point cloud processing and compression, visual information analysis, multimedia systems and applications, etc.
\end{IEEEbiography}

\end{document}